\documentclass[twocolumn,showkeys,showpacs,preprintnumbers,prd,superscriptaddress,nofootinbib]{revtex4-1}
\usepackage{graphicx,epsf,bm,amsmath,amsfonts,amssymb,epstopdf,natbib,hyperref,color,verbatim,multirow,bm,soul,physics,appendix,diagbox}
\usepackage[dvipsnames,table]{xcolor}
\usepackage[capitalize]{cleveref}
\hypersetup{colorlinks=true,urlcolor=blue,citecolor=blue,linkcolor=blue,menucolor=blue,anchorcolor=blue,filecolor=blue}

\newcolumntype{?}{!{\vrule width 3pt}}
\makeatletter
\newcommand{\fmarki}{\ensuremath{\alpha}}
\newcommand{\fmarkii}{\ensuremath{\beta}}
\newcommand{\fmarkiii}{\ensuremath{\gamma}}
\newcommand{\fmarkiv}{\ensuremath{\delta}}
\newcommand{\fmarkv}{\ensuremath{\epsilon}}
                
\def\@fnsymbol#1{{\ifcase#1\or \fmarki\or \fmarkii\or \fmarkiii\or \fmarkiv\or \fmarkv\or \fmarkvi\or \fmarkvii\or \fmarkviii\or \fmarkix\or \fmarkx\or \fmarkxi\or \fmarkxii\or \fmarkxiii\or \fmarkxiv\or \fmarkxv\or \else\@ctrerr\fi}}
\makeatother

\date{\today}
\begin{document}

\title{Combining pre- and post-recombination new physics to address cosmological tensions: case study with varying electron mass and sign-switching cosmological constant}

\author{Yo Toda}
\email{y-toda@particle.sci.hokudai.ac.jp}
\affiliation{Department of Physics, Hokkaido University, Kita 10, Nishi 8, Kita-ku, Sapporo 060-0810, Japan \looseness=-1}

\author{William Giar\`{e}}
\email{w.giare@sheffield.ac.uk}
\affiliation{School of Mathematics and Statistics, University of Sheffield, Hounsfield Road, Sheffield S3 7RH, United Kingdom \looseness=-1}

\author{Emre \"{O}z\"{u}lker}
\email{ozulker17@itu.edu.tr}
\affiliation{Department of Physics, Istanbul Technical University, 34469 Maslak, Istanbul, Turkey \looseness=-1}

\author{Eleonora Di Valentino}
\email{e.divalentino@sheffield.ac.uk}
\affiliation{School of Mathematics and Statistics, University of Sheffield, Hounsfield Road, Sheffield S3 7RH, United Kingdom \looseness=-1}

\author{Sunny Vagnozzi}
\email{sunny.vagnozzi@unitn.it}
\affiliation{Department of Physics, University of Trento, Via Sommarive 14, 38123 Povo (TN), Italy}
\affiliation{Trento Institute for Fundamental Physics and Applications (TIFPA)-INFN, Via Sommarive 14, 38123 Povo (TN), Italy}

\begin{abstract}
\noindent It has recently been argued that the Hubble tension may call for a combination of both pre- and post-recombination new physics. Motivated by these considerations, we provide one of the first concrete case studies aimed at constructing such a viable combination. We consider models that have individually worked best on either end of recombination so far: a spatially uniform time-varying electron mass leading to earlier recombination (also adding non-zero spatial curvature), and a sign-switching cosmological constant inducing an AdS-to-dS transition within the $\Lambda_{\rm s}$CDM model. When confronted against Cosmic Microwave Background (CMB), Baryon Acoustic Oscillations, and Type Ia Supernovae data, we show that no combination of these ingredients can successfully solve the Hubble tension. We find that the matter density parameter $\Omega_m$ plays a critical role, driving important physical scales in opposite directions: the AdS-to-dS transition requires a larger $\Omega_m$ to maintain the CMB acoustic scale fixed, whereas the varying electron mass requires a smaller $\Omega_m$ to maintain the redshift of matter-radiation equality fixed. Despite the overall failure, we use our results to draw general model-building lessons, highlighting the importance of assessing tension-solving directions in the parameter space of new physics parameters and how these correlate with shifts in other standard parameters, while underscoring the crucial role of $\Omega_m$ in this sense.

\end{abstract}
\preprint{EPHOU-24-007}

\maketitle

\section{Introduction}
\label{sec:introduction}

The $\Lambda$CDM model has proven to be extremely successful in accounting for a wide range of cosmological and astrophysical observations~\cite{DES:2017qwj,Planck:2018vyg,ACT:2020gnv,eBOSS:2020yzd,Brout:2022vxf}. As the precision of our instruments improves, so does that of the inferred cosmological parameters, potentially leading to mismatches between $\Lambda$CDM-based model predictions for such parameters and direct measurements thereof: the most prominent mismatch concerns the Hubble constant $H_0$.
The latest observations of Cosmic Microwave Background (CMB) temperature, polarization, and lensing from the \textit{Planck} satellite, when analyzed assuming the $\Lambda$CDM model, yield $H_0 = (67.36 \pm 0.54)\,\text{km/s/Mpc}$~\cite{Planck:2018vyg}. On the other hand, the local distance ladder measurement of $H_0$ via Cepheid-calibrated Type Ia Supernovae (SNeIa) from the SH0ES team in Ref.~\cite{Riess:2021jrx} yields $H_0 = (73.04 \pm 1.04)\,\text{km/s/Mpc}$, while a number of other local observations (see for example Refs.~\cite{Dhawan:2017ywl,LIGOScientific:2017adf,Freedman:2019jwv,Huang:2019yhh,Pesce:2020xfe,deJaeger:2020zpb,Schombert:2020pxm,Blakeslee:2021rqi,Dhawan:2022yws,YoungSupernovaExperiment:2022fqk,Tully:2022rbj,Lenart:2022nip,TDCOSMO:2023hni,Dainotti:2023ebr,Bargiacchi:2023jse,Scolnic:2023mrv,Uddin:2023iob,Dhungana:2023dep,Tully:2023epf,DESI:2023fij,Huang:2023frr,Li:2024yoe,Pascale:2024qjr}) are also consistent with higher values of $H_0$, albeit not at the same level of tension with \textit{Planck} as the SH0ES measurement. The resulting $\gtrsim 5\sigma$ Hubble tension is leading to a potential cosmological crisis, but at the same time constitutes one of the most exciting and fast-moving open problems in the field -- see e.g.\ Refs.~\cite{Verde:2019ivm,DiValentino:2020zio,DiValentino:2021izs,Perivolaropoulos:2021jda,Schoneberg:2021qvd,Shah:2021onj,Abdalla:2022yfr,DiValentino:2022fjm,Hu:2023jqc,Verde:2023lmm} for reviews. 

With the continuous improvement in precision of cosmological measurements, the Hubble tension has not only persisted but increased in significance, while a number of other milder tensions have appeared (e.g.\ the discrepancy in $S_8$, measuring the amplitude of clustering in the late Universe~\cite{DiValentino:2018gcu,DiValentino:2020vvd,Nunes:2021ipq,deSa:2022hsh,Pedreira:2023qqt}). The steady growth of the Hubble tension and the increasing difficulty in attributing it to systematics~\cite{Mortsell:2021nzg,Mortsell:2021tcx,Freedman:2021ahq,Kenworthy:2022jdh,Wojtak:2022bct,Riess:2022mme,Murakami:2023xuy,Riess:2023bfx,Riess:2024ohe,Breuval:2024lsv} has motivated within the community over a decade of attempts to build new physics models aimed at addressing this and potentially other observational tensions:\footnote{To the best of our knowledge one of the earliest works recognizing the problem, explicitly referring to it as a tension, and attempting solutions in terms of new physics was Ref.~\cite{Verde:2013wza}, which dates back to the first half of 2013.} with no claims as to completeness, see e.g.\ Refs.~\cite{Anchordoqui:2015lqa,Karwal:2016vyq,Benetti:2017juy,Mortsell:2018mfj,Kumar:2018yhh,Guo:2018ans,Graef:2018fzu,Agrawal:2019lmo,Escudero:2019gvw,Niedermann:2019olb,Sakstein:2019fmf,Ballesteros:2020sik,Jedamzik:2020krr,Ballardini:2020iws,DiValentino:2020evt,Niedermann:2020dwg,Gonzalez:2020fdy,Braglia:2020auw,RoyChoudhury:2020dmd,Brinckmann:2020bcn,Karwal:2021vpk,Gomez-Valent:2021cbe,Cyr-Racine:2021oal,Niedermann:2021ijp,Saridakis:2021xqy,Herold:2021ksg,Odintsov:2022eqm,Aboubrahim:2022gjb,Ren:2022aeo,Adhikari:2022moo,Nojiri:2022ski,Schoneberg:2022grr,Joseph:2022jsf,Gomez-Valent:2022bku,Odintsov:2022umu,Ge:2022qws,Schiavone:2022wvq,Brinckmann:2022ajr,Khodadi:2023ezj,Kumar:2023bqj,Ben-Dayan:2023rgt,Ruchika:2023ugh,Yadav:2023yyb,Sharma:2023kzr,Ramadan:2023ivw,Fu:2023tfo,Efstathiou:2023fbn,Montani:2023ywn,Stahl:2024stz,Garny:2024ums,Co:2024oek} for examples of these models. However, to better appreciate at what epochs new physics may be required to resolve the $H_0$ tension, it is useful to inspect the role of the comoving size of the sound horizon at the epoch of recombination, $r_s$. Datasets calibrated with early-time (pre-recombination) information appear to be consistent with the low value of $H_0$ inferred from \textit{Planck}: a key example in this sense are Baryon Acoustic Oscillation (BAO) measurements calibrated by using a Big Bang Nucleosynthesis (BBN) prior on the physical baryon density $\omega_b$ to determine $r_s$. Further combining BBN and BAO measurements with uncalibrated (Hubble flow) SNeIa allows one to build an \textit{inverse distance ladder}, from which a low value of $H_0$ consistent with the \textit{Planck} $\Lambda$CDM one, yet \textit{completely independent of any CMB data}, is inferred. These results are in stark disagreement with the higher values of $H_0$ obtained via local distance measurements which calibrate SNeIa (or, more generally, other distance indicators) by inferring their absolute magnitude, $M_B$, utilizing nearby astrophysical objects. The tension between local and inverse distance ladder inferences of $H_0$ indicates that, at its heart, the Hubble tension is actually a tension between calibrators, which can be cleanly recast in terms of the SNeIa absolute magnitude $M_B$, and the sound horizon $r_s$~\cite{Bousis:2024rnb}.

Since BAO data on their own are sensitive to the combination $H_0r_s$,\footnote{BAO measurements are actually sensitive to the sound horizon evaluated at the baryon drag epoch, typically denoted by $r_{\rm d}$, whereas the scale governing the physics behind the acoustic peaks in the CMB is the sound horizon evaluated at recombination, typically denoted by $r_{\star}$. The two epochs are separated in redshift by $\Delta z = z_d - z_{\star} \sim 30$, so the difference between the two sound horizons is very small, but nevertheless important. In order not to overburden the notation, here we shall simply refer to the ``sound horizon'' denoted by $r_s$, and it will be clear from the context of the discussion which of the two sound horizons we are actually referring to.} it is clear that (assuming $M_B$ is left untouched) the only way to get local and inverse distance ladder measurements of $H_0$ to agree is to lower $r_s$ -- given that the integral determining $r_s$ runs from the earliest times up to recombination, this necessarily requires introducing new physics prior to recombination~\cite{Bernal:2016gxb,Addison:2017fdm,Lemos:2018smw,Aylor:2018drw,Schoneberg:2019wmt,Knox:2019rjx,Arendse:2019hev,Efstathiou:2021ocp,Cai:2021weh,Keeley:2022ojz}. One interesting possibility in this sense is that of increasing the pre-recombination expansion rate, for instance through an early dark energy (EDE) component~\cite{Poulin:2018cxd,Kamionkowski:2022pkx,Poulin:2023lkg}. Another interesting class of proposals invokes new physics which makes recombination occur earlier, for instance via a time-varying electron mass $m_e$~\cite{Hart:2019dxi,Sekiguchi:2020teg}, as we will discuss in significantly more detail later.

The above considerations make it clear that it is extremely difficult, if not impossible, to resolve the Hubble tension by invoking exclusively late-time, post-recombination physics~\cite{Bernal:2016gxb,Addison:2017fdm,Lemos:2018smw,Aylor:2018drw,Schoneberg:2019wmt,Knox:2019rjx,Arendse:2019hev,Efstathiou:2021ocp,Cai:2021weh,Keeley:2022ojz}. Because they are on their own unable to alter the sound horizon, late-time modifications to $\Lambda$CDM in the presence of a higher value of $H_0$ are bound to worsen the fit to BAO and uncalibrated SNeIa data, which control respectively the normalization and shape of the $z \lesssim 2$ expansion rate.\footnote{This argument does not necessarily hold, or is at the very least weakened, if one considers 2D BAO data, which have been argued to carry less cosmological model-dependence compared to their widely used 3D counterparts (see e.g.\ Refs.~\cite{Carvalho:2015ica,deCarvalho:2017xye,Carvalho:2017tuu,Camarena:2019rmj,Nunes:2020hzy,Nunes:2020uex,deCarvalho:2021azj,Staicova:2021ntm,Menote:2021jaq,Benisty:2022psx,Bernui:2023byc,Shah:2024slr,Favale:2024sdq,Ruchika:2024lgi,Giare:2024syw} for examples of discussions on 2D BAO data and their use in obtaining cosmological constraints). Nevertheless, in order to be as conservative as possible, we will not use this less conventional approach in the present work. In what follows, BAO will thus always refer to 3D BAO measurements.} Nevertheless, some amount of late-time new physics which can partially (but not completely) alleviate the Hubble tension is allowed by BAO and SNeIa data, and various proposals in this direction have been explored in the literature, mostly in relation to new dynamics in the dark energy (DE) sector (see e.g.\ Refs.~\cite{Zhao:2017urm,DiValentino:2017iww,Vagnozzi:2018jhn,Yang:2018euj,Banihashemi:2018oxo,Banihashemi:2018has,Li:2019yem,Yang:2019nhz,Vagnozzi:2019ezj,Hogg:2020rdp,Alestas:2020mvb,Banerjee:2020xcn,DiValentino:2020kha,Banihashemi:2020wtb,Alestas:2020zol,Gao:2021xnk,Alestas:2021xes,Heisenberg:2022lob,Heisenberg:2022gqk,Sharma:2022ifr,Nunes:2022bhn,Sharma:2022oxh,Moshafi:2022mva,Banerjee:2022ynv,Alvarez:2022wef,Gangopadhyay:2022bsh,Gao:2022ahg,Dahmani:2023bsb,deCruzPerez:2023wzd,Ballardini:2023mzm,Yao:2023ybs,Gangopadhyay:2023nli,Zhai:2023yny,SolaPeracaula:2023swx,Gomez-Valent:2023hov,Frion:2023xwq,Escamilla:2023oce,Hoerning:2023hks,Petronikolou:2023cwu,Ben-Dayan:2023htq,Lazkoz:2023oqc,Forconi:2023hsj,Sebastianutti:2023dbt,Benisty:2024lmj,Giare:2024ytc,Shah:2024rme,Giare:2024smz,Montani:2024pou,Jia:2024wix,Aboubrahim:2024spa}).

Despite a plethora of attempts, it is fair to say that no compelling model of early-time new physics which is \textit{i}) able to fully solve the Hubble tension while \textit{ii}) maintaining a good fit to all available cosmological data \textit{iii}) without worsening other milder observational discrepancies, has yet been constructed. Empirically speaking, it is the case that even the most successful models appear, at best, to bring the value of $H_0$ inferred from CMB, BAO, and Hubble flow SNeIa data (therefore without an inclusion of SH0ES $H_0$ prior) up to $H_0 \sim 70\,\text{km/s/Mpc}$. Could it perhaps be that models of early-time new physics, on their own, may not be enough to fully solve the Hubble tension? This conclusion was recently put forward by one of us in Ref.~\cite{Vagnozzi:2023nrq}, on the basis of seven assorted hints~\cite{Vagnozzi:2021tjv,Jedamzik:2020zmd,Wong:2019kwg,Krishnan:2020obg,Krishnan:2020vaf,Dainotti:2021pqg,Dainotti:2022bzg,Colgain:2022tql,Jia:2022ycc,Vagnozzi:2021gjh,Lin:2019htv,Lin:2021sfs,Baxter:2020qlr,Philcox:2020xbv,Philcox:2022sgj,Akarsu:2024qiq}: in particular, it was argued that solving the Hubble tension will ultimately require a combination of early- and late-time new physics, and potentially local new physics.\footnote{For studies on the impact of local or very late-time physics on the Hubble tension, see e.g.\ Refs.~\cite{Kenworthy:2019qwq,Desmond:2019ygn,Ding:2019mmw,Benevento:2020fev,Desmond:2020wep,Cai:2020tpy,Camarena:2021jlr,Cai:2021wgv,Marra:2021fvf,Krishnan:2021dyb,Perivolaropoulos:2021bds,Castello:2021uad,Camarena:2022iae,Perivolaropoulos:2022khd,Oikonomou:2022tjm,Hogas:2023vim,Hogas:2023pjz,Giani:2023aor,Mazurenko:2023sex,Huang:2024erq}.} However, besides a few general considerations, Ref.~\cite{Vagnozzi:2023nrq} did not propose any concrete combination of potentially interesting early-plus-late-time new physics models, delegating the quest of finding one such combination as an exercise to the reader.

The goal of the present work is precisely that of embracing the above challenge, thereby providing one of the first concrete case studies aimed at combining early- and late-time new physics in an attempt to address the Hubble tension (see also Refs.~\cite{Allali:2021azp,Anchordoqui:2021gji,Khosravi:2021csn,Clark:2021hlo,Wang:2022jpo,Anchordoqui:2022gmw,Reeves:2022aoi,Yao:2023qve,daCosta:2023mow,Wang:2024dka,Baryakhtar:2024rky} for earlier studies which in part show a similar spirit, featuring combinations of modifications and/or ingredients at early and late times). The choice of which specific models to combine is guided by considerations on which models have \textit{individually} shown the most promise so far. In this first work, our (admittedly na\"{i}ve) zeroth-order criterion to judge promise, therefore, is simply assessing whether the value of $H_0$ inferred within the models in question from CMB+BAO+SNeIa data (without a SH0ES $H_0$ prior) is sufficiently high. More specifically, on the early-time side we consider the $\Lambda$CDM$+m_e$ model, featuring a spatially uniform time-varying electron mass $m_e$~\cite{Hart:2019dxi,Sekiguchi:2020teg}, and introducing an extra parameter, i.e.\ the fractional variation in $m_e$ between recombination and today: a higher value of $m_e$ leads to recombination occurring earlier, thereby reducing $r_s$.
Once the curvature parameter $\Omega_K$ is also allowed to vary ($\Lambda$CDM$+m_e+\Omega_K$ model), as required to provide a good fit to late-time data~\cite{Sekiguchi:2020teg}, the resulting model emerges as one of the most promising early-time modifications, as explicitly quantified by means of the metrics introduced in Ref.~\cite{Schoneberg:2021qvd}. On the late-time side, we instead consider the so-called $\Lambda_{\rm s}$CDM model~\cite{Akarsu:2021fol}, which introduces a rapid sign switch in the cosmological constant (from $-\Lambda$ to $\Lambda$) at redshifts $z_{\dagger} \sim 2$. While on its own the model cannot completely solve the Hubble tension, it can nonetheless alleviate it to an interesting extent while also addressing a number of other discrepancies. In the spirit of the proposal of Ref.~\cite{Vagnozzi:2023nrq}, we therefore consider various combinations of time-varying electron mass, non-zero spatial curvature, and sign-switching cosmological constant: in short, our goal is to assess whether the whole is greater than the sum of its parts, i.e.\ whether the combination of these ingredients performs better than the individual ingredients themselves in addressing the Hubble tension. While we do not find this to be the case, at least for the specific combination of ingredients considered, our analysis teaches us a number of valuable lessons and considerations (especially regarding the matter density parameter $\Omega_m$) for future attempts at solving the Hubble tension by combining early- and late-time new physics, a quest which turns out to be much more difficult than expected (partly due to reasons anticipated in Ref.~\cite{Vagnozzi:2023nrq}).

The rest of this paper is then organized as follows. In Sec.~\ref{sec:models} we review in more detail the two early- and late-time new physics models we combine: $\Lambda$CDM$+m_e$ ($+\Omega_K$) and $\Lambda_{\rm s}$CDM. In Sec.~\ref{sec:data} we discuss the datasets and methodology we make use of. The results of our analysis can be found in Sec.~\ref{sec:results}. Finally, in Sec.~\ref{sec:conclusions} we draw a number of concluding remarks and summarize the lessons learned from our case study.

\section{Models}
\label{sec:models}

In what follows, we discuss in more detail the two models of early- and late-time new physics we combine: a spatially uniform time-varying electron mass (potentially in a non-spatially flat Universe) and a sign-switching cosmological constant.

\subsection{Varying electron mass in a non-flat Universe}
\label{subsec:lcdmme}

For what concerns early-time new physics, we consider a model where the value of the electron mass in the early Universe $m_e$ differs from its present value $m_{e,0} \simeq 511\,{\text{keV}}$; we refer to this model as $\Lambda$CDM$+m_e$. Indeed, a variation in the fundamental properties of Hydrogen and Helium is one of the most effective ways of shifting the epoch of recombination, and thereby the sound horizon at recombination which, as argued in Sec.~\ref{sec:introduction}, plays a key role in the Hubble tension discussion. The cosmological implications of a varying electron mass have been discussed in a number of earlier works, including Refs.~\cite{Planck:2014ylh,Chluba:2015gta,Hart:2017ndk,Khalife:2023qbu,Chluba:2023xqj}. In particular, varying $m_e$ alters a number of relevant quantities, including most importantly the energy levels of Hydrogen and Helium, as well as the Thomson scattering cross section, two-photon decay rate of the second shell, effective recombination and photoionization rates, effective temperature at which the former are evaluated, effective dipole transition rate for the Lyman-$\alpha$ resonance, K-factors, Einstein A coefficients, and so on. We refer the reader to Refs.~\cite{Planck:2014ylh,Hart:2017ndk} for recent more detailed discussions on these and other effects.

For our purposes, it is the effects of a change in $m_e$ on the Hydrogen atom energy levels and Thomson scattering cross section which are most relevant in terms of cosmological implications. In particular, the energy levels scale as $E_i \propto m_e$, whereas the Thomson scattering cross-section scales as $\sigma_T \propto m_e^{-2}$ (the latter is nothing other than a consequence of the behavior of the electron propagator in the non-relativistic regime, where $p^2 \ll m_e^2$). Focusing on the effects on $E_i$, this implies that an increase in the electron mass at recombination raises the binding energy of neutral Hydrogen. As a consequence, recombination becomes energetically favorable at higher energies and thereby occurs earlier, reducing the redshift range over which the integral determining the sound horizon is evaluated, and therefore reducing the sound horizon $r_s$ itself. As discussed in Sec.~\ref{sec:introduction}, this is a key feature in order for a cosmological model to accommodate a higher $H_0$ without ruining the fit to BAO data.

However, the scaling of $\sigma_T$ with $m_e$ is also crucial for the model to succeed in leading to higher values of $H_0$. A key challenge faced by many models of early-time new physics, particularly those increasing the pre-recombination expansion rate $H(z)$, is that they alter the ratio $\theta_d/\theta_s$, which is exquisitely constrained from CMB measurements: this is the ratio between the angular sizes on the last-scattering surface of the diffusion length at last scattering $r_d$ (basically the ``Silk damping scale,'' not to be confused with the sound horizon at the drag epoch, $r_{\rm d}$) and the sound horizon at last scattering $r_s$. The reason behind this has been discussed in detail in Ref.~\cite{Knox:2019rjx} and can be understood as follows: early-time new physics leads to a fractional reduction in the sound horizon $\delta r_s/r_s$, while also leading to an identical fractional reduction in the angular diameter distance to last-scattering $d_A$, $\delta d_A/d_A \approx \delta r_s/r_s$, in order to keep $\theta_s$ fixed. However, for a large class of early-time new physics models (the prototype being an increase in the effective number of relativistic species, $N_{\text{eff}}$), the fractional change in the diffusion length is $\delta r_d/r_d \approx 1/2 \delta r_s/r_s$,\footnote{The factor of $1/2$ is ultimately tied to the random walk nature of diffusion damping, which results in the expression for $r_d$ having a similar functional form to that for $r_s$, but with a square root in front. This can also be understood by looking at Fig.~2 of Ref.~\cite{Knox:2019rjx}, where it is clear that the kernel describing the fractional linear response of $r_d$ to a fractional change in $H(z)$ is centered over a much narrower redshift range compared to the analogous quantity for $r_s$.} which implies that $\theta_d$ necessarily changes if $\theta_s$ is kept fixed, thereby altering $\theta_d/\theta_s$: this is undesirable and limits the extent to which CMB data allows for such early-time new physics.
However, the scaling $\sigma_T \propto m_e^{-2}$ improves the situation, since the integrand of the integral determining $r_d$ scales as $1/\sqrt{\sigma_T}$. This allows for additional contributions to $\delta r_d/r_d$, which help in maintaining $\theta_d/\theta_s$ fixed, thereby improving agreement with CMB observations on small scales. The crucial role played by the $\sigma_T \propto m_e^{-2}$ scaling was explicitly noted by Refs.~\cite{Hart:2017ndk,Hart:2019dxi}, where it was shown that properly including this scaling and its effects on the Thomson visibility function opens up a large geometrical degeneracy between $m_e$ and $H_0$, allowing for significantly larger values of $H_0$ compared to the case where the Thomson scattering cross section is not properly scaled when $m_e$ is varied.

From the above considerations, it is now clear why a model where $m_e$ was a few percent larger at recombination compared to its present-day value, $m_{e,0}$, can provide a potential solution to the Hubble tension, as studied in a number of recent works (see e.g., Refs.~\cite{Hart:2019dxi,Schoneberg:2021qvd,Seto:2022xgx,Hoshiya:2022ady,Solomon:2022qqf,Seto:2024cgo}). However, as explicitly noted in Ref.~\cite{Sekiguchi:2020teg}, a model in which only $m_e$ is varied while assuming a spatially flat $\Lambda$CDM background at late times cannot accommodate an adequately large transition in the electron mass so as to shrink the sound horizon to small enough values. The reason is that CMB data imposes strong correlations between the baryonic and dark matter densities and the time of recombination (which is modified by the altered electron mass) -- these correlations cause the baryonic and dark matter density parameters to deviate from their usual $\Lambda$CDM values by a far margin when the sound horizon is significantly reduced. These induced parameter shifts spoil the good fit of the late-time $\Lambda$CDM cosmology to BAO and SNeIa distance measurements. Nevertheless, it has been shown that allowing for non-zero spatial curvature allows one to overcome this problem and thereby accommodate a higher $H_0$ while increasing $m_e$, while simultaneously delivering a good fit to CMB, BAO, and SNeIa data. As shown in Ref.~\cite{Sekiguchi:2020teg}, this requires a negative spatial curvature parameter $\Omega_K<0$, which corresponds to a spatially closed Universe.\footnote{This is of particular interest in light of recent discussions on the possibility that \textit{Planck} data may favor a spatially closed Universe~\cite{DiValentino:2019qzk,Handley:2019tkm}. Without seeking to take sides in this discussion, we refer the reader for instance to Refs.~\cite{Efstathiou:2020wem,DiValentino:2020hov,Benisty:2020otr,Vagnozzi:2020rcz,Vagnozzi:2020dfn,DiValentino:2020kpf,Yang:2021hxg,Cao:2021ldv,Dhawan:2021mel,Gonzalez:2021ojp,Dinda:2021ffa,Zuckerman:2021kgm,Bargiacchi:2021hdp,Glanville:2022xes,Bel:2022iuf,Yang:2022kho,Stevens:2022evv,Favale:2023lnp,Qi:2023oxv,Giare:2023ejv} for recent studies on various aspects of this issue, and possible ways to arbitrate it. We note that the question is far from being settled, and exploring the role spatial curvature may play in the Hubble tension is therefore still of interest.}

In light of the above considerations, we envisage a class of models featuring a spatially uniform variation of $m_e$. In particular, as done in earlier works~\cite{Hart:2019dxi,Sekiguchi:2020teg,Schoneberg:2021qvd}, $m_e(z)$ is approximated as being redshift-independent over the narrow redshift range of recombination, whereas its full functional form after recombination is irrelevant to cosmological observables as long as the present-day value $m_{e,0}$ is reached at a reasonably high redshift. Therefore, this amounts to introducing the extra parameter $m_e/m_{e,0}$, where by $m_e$ we denote the value of the electron mass at recombination. With the above discussions in mind, we consider different cosmological models where either or both $m_e/m_{e,0}$ and $\Omega_K$ are varied. Note that, because of the necessity of varying $\Omega_K$ to successfully fit all available data within the varying electron mass model, the combined model, $\Lambda$CDM$+m_e+\Omega_K$, is strictly speaking not a purely early-time new physics model. Nevertheless, for simplicity, we place it within this category, as usually done in the literature. Finally, we note that from the model-building point of view, models with varying fundamental constants (such as the electron mass) can be embedded within theories featuring direct couplings of the matter fields to additional scalar fields, such as a dilaton or the dark energy field~\cite{Carroll:1998zi,Brax:2002nt,Chiba:2006xx,Baryakhtar:2024rky}.

\subsection{Sign-switching cosmological constant: the $\Lambda_{\rm s}$CDM model}
\label{subsec:lscdm}

The $\Lambda_{\rm s}$CDM model~\cite{Akarsu:2021fol} is an extension of the baseline $\Lambda$CDM model that replaces the usual cosmological constant with a sign-switching one, $\Lambda_{\rm s}$, which at the transition redshift $z=z_{\dagger}$ abruptly shifts its value from $\Lambda_{\rm AdS}<0$ to $\Lambda_{\rm dS}=-\Lambda_{\rm AdS}$, with the latter then driving the late-time acceleration of the Universe. Therefore, the model is a phenomenological implementation of an abrupt anti-de Sitter (AdS) to de Sitter (dS) vacuum transition. The cosmological and astrophysical implications of this model was investigated in Refs.~\cite{Akarsu:2021fol,Akarsu:2022typ,Akarsu:2023mfb,Paraskevas:2023itu,Akarsu:2024eoo,Paraskevas:2024ytz,Yadav:2024duq}. The precursor of the $\Lambda_{\rm s}$CDM model, from which the latter took inspiration, is the so-called graduated DE model. First studied in Ref.~\cite{Akarsu:2019hmw}, this model dynamically modifies the functional form of the null inertial mass density of the vacuum energy, while also promoting its value (equal to zero in the standard case) to a negative constant. The resulting model, when constrained against cosmological data, was shown to effectively lead to a very rapid sign-switch in the cosmological constant, while at the same time partially relaxing (but not completely solving) the $H_0$ tension~\cite{Akarsu:2019hmw}.

Inspired by the above findings, Ref.~\cite{Akarsu:2021fol} proposed the $\Lambda_{\rm s}$CDM model, where the gradual transition within the graduated DE model is replaced by a phenomenological, abrupt AdS-to-dS vacuum transition at $z_{\dagger} \sim 2$:
\begin{eqnarray}
\Lambda \,\,\rightarrow\,\, \Lambda_{\rm s} \equiv \Lambda_{\rm s0}\,{\text{sgn}}[z_{\dagger}-z]\,,
\label{eq:lambda}
\end{eqnarray}
where $\Lambda_{\rm s0}$ is the present-day value of the positive cosmological constant, and the signum function is given by:
\begin{eqnarray}
{\text{sgn}}[x] =
\begin{cases}
-1 \quad \quad &(x<0)\,, \\
0 \quad \quad &(x=0)\,, \\
+1 \quad \quad &(x>0)\,. \\
\end{cases}
\label{eq:sgn}
\end{eqnarray}
This model is typically categorized as a late-time modification to $\Lambda$CDM even though, strictly speaking, the AdS phase persists even at earlier times, including before recombination. However, just as the positive cosmological constant of $\Lambda$CDM is completely negligible during and prior to recombination, the same is true for the negative cosmological constant of $\Lambda_{\rm s}$CDM, given that the two have the same absolute value. In fact, the negative cosmological constant leads to a fractional contribution of $\lesssim 10^{-9}$ to the energy budget of the Universe at recombination, and has therefore virtually no effect on quantities related to early-Universe physics (including, most importantly, the sound horizon $r_s$). Note that the $\Lambda$CDM model is recovered only in the limit where $z_{\dagger} \to \infty$.

Initially introduced as a phenomenological model for cosmology, recent investigations have looked into potential microphysical origins for the $\Lambda_{\rm s}$CDM model, which include but are not limited to sign switch from Casimir forces~\cite{Anchordoqui:2023woo} and type-II minimally modified theories of gravity~\cite{Akarsu:2024qsi,Akarsu:2024eoo}.\footnote{It is worth mentioning that the $\Lambda_{\rm s}$CDM model embedded in these type-II minimally modified theories of gravity appears slightly more promising than the fully phenomenological model as the embedding provides a theoretical basis for the sign-switch, and its explicit treatment of the perturbations during the transition period (in contrast with the phenomenological model where the transition is assumed to be instantaneous with no effect on the perturbations) appears to lead to the model parameters being better constrained with CMB-only data, while also yielding slightly higher inferred values for $H_0$~\cite{Akarsu:2024eoo}.} It is worth noting that the possibility of cosmologies with negative energy densities is not a far-fetched one, but on the contrary has been the subject of a number of studies in recent years, not limited to the $\Lambda_{\rm s}$CDM model (see for example Refs.~\cite{Poulin:2018zxs,Dutta:2018vmq,Visinelli:2019qqu,Haridasu:2019gyz,Ruchika:2020avj,DiValentino:2020naf,Calderon:2020hoc,Hogas:2021fmr,Sen:2021wld,Ong:2022wrs,Ozulker:2022slu,Malekjani:2023ple,VanRaamsdonk:2023ion,Adil:2023exv,Adil:2023ara,Gomez-Valent:2023uof,Menci:2024rbq,Gomez-Valent:2024tdb,DESI:2024aqx,Colgain:2024clf,VanRaamsdonk:2024sdp,Wang:2024hwd} in the context of late-time DE, Refs.~\cite{Ye:2020btb,Ye:2020oix,Ye:2021nej,Jiang:2021bab,Ye:2021iwa,Jiang:2022qlj,Jiang:2022uyg,Jiang:2023bsz,Ye:2023zel,Peng:2023bik} in the context of EDE, and Refs.~\cite{Craig:2024tky,Noriega:2024lzo,Green:2024xbb,Elbers:2024sha,Naredo-Tuero:2024sgf} in the context of effective negative neutrino masses). Indeed, the possibility of negative energy densities enjoys strong motivation from string theory, wherein stable dS vacua have proven extremely difficult to construct, but AdS vacua are ubiquitous: it has in fact been conjectured that string theory may be unable to harbor dS vacua, as advocated by the swampland program~\cite{Vafa:2005ui,Danielsson:2018ztv,Palti:2019pca}, with important cosmological consequences, particularly for the accelerated phases of the cosmological expansion~\cite{Obied:2018sgi,Agrawal:2018own,Achucarro:2018vey,Garg:2018reu,Kehagias:2018uem,Kinney:2018nny,Ooguri:2018wrx,Odintsov:2020zkl,Oikonomou:2020oex,Cicoli:2021skd}.

If the abrupt jump is taken at face value, it corresponds to a non-differentiable scale factor $a(t)$, alongside a discontinuous $\dot{a}(t)$, and a singular $\ddot{a}(t)$, implying a type II (sudden) singularity~\cite{Barrow:2004xh,deHaro:2023lbq,Trivedi:2023zlf}. Nevertheless, this idealized instantaneous transition captures the essence of $\Lambda_{\rm s}$CDM very well and is convenient for numerical studies thanks to its simplicity. However, should one wish to work with a more realistic parametrization of a smooth $\Lambda$ term incorporating a more gradual transition, other functions such as sigmoids can be utilized, e.g.\ $\Lambda_{\rm s}(z) = \Lambda_{\rm s0} \tanh[\eta(z_\dagger-z)]/\tanh[\eta z_\dagger]$ as proposed in Refs.~\cite{Akarsu:2022typ,Akarsu:2024qsi}. Throughout this work, unless otherwise specified, we shall work with the instantaneous transition described by Eq.~(\ref{eq:lambda}). Some astrophysical consequences of such an abrupt transition, particularly for bound cosmic structures (e.g.\ virialized galaxy clusters), have been recently studied in Refs.~\cite{Paraskevas:2023itu,Paraskevas:2024ytz}: these findings show that the abrupt transition does not result in the dissociation of bound systems, but rather exerts relatively weak effects thereon, further reinforcing the viability of the sudden transition parametrization.

The $\Lambda_{\rm s}$CDM model has demonstrated particular promise in alleviating (albeit not completely solving, for reasons which are now well understood and discussed in Sec.~\ref{sec:introduction}) the Hubble tension, and a number of other less statistically significant cosmological discrepancies, including the $S_8$ tension, the BAO Lyman-$\alpha$ discrepancy, the cosmic age issue, and so on~\cite{Akarsu:2021fol,Akarsu:2022typ,Akarsu:2023mfb,Akarsu:2024eoo}. The reason why the $\Lambda_{\rm s}$CDM model is able to partially alleviate the Hubble tension is because its lower energy density compared to $\Lambda$CDM during the AdS phase ($z>z_{\dagger}$) would increase the distance to the last scattering surface, thereby reducing $\theta_s$ since the sound horizon is not affected; as the inferred $\theta_s$ from CMB measurement is extremely robust, such a change is not allowed and needs to be compensated by increasing $H_0$. The mechanism is similar to that by which phantom DE models (with DE equation of state $w<-1$) lead to higher values of $H_0$, the common feature being a DE density that decreases towards the past, reducing the total energy density compared to the $\Lambda$CDM model -- yet, compared to these models, $\Lambda_{\rm s}$CDM performs better for two reasons. Firstly, its DE density takes negative values, requiring a larger $H_0$ value to compensate the shift in $\theta_s$. Phantom DE models cannot transition to negative values trivially, as the sign change of the energy density of a minimally coupled source requires a singular equation of state parameter~\cite{Ozulker:2022slu}. However, this can be achieved by combining phantom DE models with an AdS vacuum as was done in Refs.~\cite{Visinelli:2019qqu,Adil:2023ara,Menci:2024rbq}. Secondly, the $\Lambda_{\rm s}$CDM model is able to offer a better fit to BAO and SNeIa data since the shape of the expansion history is simply that of $\Lambda$CDM for $z<z_{\dagger}$ as preferred by these late-time data, albeit with eventually different values of the common cosmological parameters. On the other hand, phantom DE models lead to a different shape of the expansion history at all redshifts, which is tightly constrained by late-time observations. In light of the above interesting features, we decide to consider the $\Lambda_{\rm s}$CDM model as our ``representative'' model for late-time new physics, alongside the non-zero spatial curvature parameter already discussed earlier in Sec.~\ref{subsec:lcdmme} and required for a working early-time new physics model based on a varying electron mass.

\section{Datasets and methodology}
\label{sec:data}

\begin{table*}[!t]
\scalebox{1.2}{
\begin{tabular}{|c?c|c|}
\hline
\textbf{Model} & \texttt{\#} parameters & Free parameters \\
\hline \hline
$\Lambda$CDM & 6 & $\omega_b$, $\omega_c$, $\theta_s$, $A_s$, $n_s$, $\tau$ \\ \hline
$\Lambda$CDM$+m_e$ & 7 & $\omega_b$, $\omega_c$, $\theta_s$, $A_s$, $n_s$, $\tau$, $\frac{m_e}{m_{e,0}}$ \\ \hline
$\Lambda$CDM$+\Omega_K$ & 7 & $\omega_b$, $\omega_c$, $\theta_s$, $A_s$, $n_s$, $\tau$, $\Omega_K$ \\ \hline
$\Lambda$CDM$+m_e+\Omega_K$ & 8 & $\omega_b$, $\omega_c$, $\theta_s$, $A_s$, $n_s$, $\tau$, $\frac{m_e}{m_{e,0}}$, $\Omega_K$ \\ \hline
$\Lambda_{\rm s}$CDM & 7 & $\omega_b$, $\omega_c$, $\theta_s$, $A_s$, $n_s$, $\tau$, $z_{\dagger}$ \\ \hline
$\Lambda_{\rm s}$CDM$+m_e$ & 8 & $\omega_b$, $\omega_c$, $\theta_s$, $A_s$, $n_s$, $\tau$, $z_{\dagger}$, $\frac{m_e}{m_{e,0}}$ \\ \hline
$\Lambda_{\rm s}$CDM$+\Omega_K$ & 8 & $\omega_b$, $\omega_c$, $\theta_s$, $A_s$, $n_s$, $\tau$, $z_{\dagger}$, $\Omega_K$ \\ \hline
$\Lambda_{\rm s}$CDM$+m_e+\Omega_K$ & 9 & $\omega_b$, $\omega_c$, $\theta_s$, $A_s$, $n_s$, $\tau$, $z_{\dagger}$, $\frac{m_e}{m_{e,0}}$, $\Omega_K$ \\ \hline
\end{tabular}}
\caption{Summary of the 8 cosmological models considered in this work.}
\label{tab:models}
\end{table*}

We recall that our goal in this work is to provide a case study of pre-plus-post-recombination new physics as a route towards solving the Hubble tension, in the spirit of Ref.~\cite{Vagnozzi:2023nrq}. With this in mind, and driven by the considerations made in Sec.~\ref{sec:models}, we consider various combinations of time-varying $m_e$, non-zero spatial curvature, and sign-switching cosmological constant. Specifically, the cosmological models we consider are the following:
\begin{itemize}
\item the 6-parameter $\boldsymbol{\Lambda}$\textbf{CDM} model, where the 6 parameters varied are the acoustic angular scale $\theta_s$, the physical baryon and cold dark matter densities $\omega_b$ and $\omega_c$, the amplitude and tilt of the primordial spectrum of scalar perturbations $A_s$ and $n_s$, and the optical depth to reionization $\tau$ -- this is our baseline model;
\item the 7-parameter $\boldsymbol{\Lambda}$\textbf{CDM+}$\boldsymbol{m_e}$ model, where in addition to the 6 $\Lambda$CDM parameters we vary $m_e/m_{e,0}$, the ratio of the electron mass at recombination to its current value -- this model therefore reduces to $\Lambda$CDM in the limit where $m_e/m_{e,0}=1$;
\item the 8-parameter $\boldsymbol{\Lambda}$\textbf{CDM+}$\boldsymbol{m_e}$\textbf{+}$\boldsymbol{\Omega_K}$ model, where in addition to the 6 $\Lambda$CDM parameters we vary $m_e/m_{e,0}$ and the spatial curvature parameter $\Omega_K$ -- this model therefore reduces to $\Lambda$CDM in the limit where $m_e/m_{e,0}=1$ and $\Omega_K=0$;
\item the 7-parameter $\boldsymbol{\Lambda_{\rm s}}$\textbf{CDM} model, where in addition to the 6 $\Lambda$CDM parameters we vary $z_{\dagger}$, the redshift at which the AdS-to-dS transition (sign-switch in the cosmological constant) takes place -- this model therefore reduces to $\Lambda$CDM in the limit where $z_{\dagger} \to \infty$;
\item the 8-parameter $\boldsymbol{\Lambda_{\rm s}}$\textbf{CDM+}$\boldsymbol{m_e}$ model, where in addition to the 6 $\Lambda$CDM parameters we vary $z_{\dagger}$ and $m_e/m_{e,0}$ -- this model therefore reduces to $\Lambda$CDM in the limit where $z_{\dagger} \to \infty$ and $m_e/m_{e,0}=1$;
\item the 9-parameter $\boldsymbol{\Lambda_{\rm s}}$\textbf{CDM+}$\boldsymbol{m_e}$\textbf{+}$\boldsymbol{\Omega_K}$ model, where in addition to the 6 $\Lambda$CDM parameters we vary $z_{\dagger}$, $m_e/m_{e,0}$, and $\Omega_K$ -- this model therefore reduces to $\Lambda$CDM in the limit where $z_{\dagger} \to \infty$, $m_e/m_{e,0}=1$, and $\Omega_K=0$;
\end{itemize}
For completeness, we also consider the following 2 cosmological models:
\begin{itemize}
\item the 7-parameter $\boldsymbol{\Lambda}$\textbf{CDM+}$\boldsymbol{\Omega_K}$ model, where in addition to the 6 $\Lambda$CDM parameters we vary $\Omega_K$ -- this model therefore reduces to $\Lambda$CDM in the limit where $\Omega_K=0$;
\item the 8-parameter $\boldsymbol{\Lambda_{\rm s}}$\textbf{CDM+}$\boldsymbol{\Omega_K}$ model, where in addition to the 6 $\Lambda$CDM parameters we vary $z_{\dagger}$ and $\Omega_K$ -- this model therefore reduces to $\Lambda$CDM in the limit where $z_{\dagger} \to \infty$ and $\Omega_K=0$;
\end{itemize}
For the convenience of the reader, the properties of these 8 models are summarized in Tab.~\ref{tab:models}. We set wide, flat priors on all the cosmological parameters listed above, verifying a posteriori that our posteriors are not affected by the choice of lower and upper prior boundaries. The only exception is $z_{\dagger}$, for which we set a prior $z_{\dagger} \in [1;3.5]$. However, our results are not expected to be sensitive to the upper edge of this prior, as we discuss later in Sec.~\ref{sec:results}.

To constrain the parameter space of the models, we consider various combinations of the following datasets:
\begin{itemize}
\item measurements of CMB temperature anisotropy and polarization power spectra, their cross-spectra, and lensing potential power spectrum reconstructed from the temperature 4-point function, as obtained from the \textit{Planck} 2018 legacy data release~\cite{Planck:2018lbu,Planck:2019nip};
\item BAO measurements from the 6dFGS, SDSS-MGS, and BOSS DR12 surveys~\cite{Beutler:2011hx, Ross:2014qpa, BOSS:2016wmc} -- note that we only use distance/expansion rate measurements, whereas we do not include growth rate measurements from redshift-space distortions;
\item uncalibrated SNeIa measurements from the \textit{Pantheon} sample~\cite{Pan-STARRS1:2017jku};
\item BAO measurements from the eBOSS Lyman-$\alpha$ (Ly$\alpha$) auto-correlation and Ly$\alpha$-quasar cross-correlation~\cite{eBOSS:2019qwo}, which we refer to as \textit{Ly$\alpha$};
\item a Gaussian prior on the Hubble constant $H_0=(73.04 \pm 1.04)\,\text{km/s/Mpc}$ as measured by the SH0ES team using a distance ladder approach via Cepheid-calibrated SNeIa in Ref.~\cite{Riess:2021jrx}, and referred to as \textit{R21}.
\end{itemize}
The combination of the first three datasets (CMB, BAO without \textit{Ly$\alpha$} measurements, and SNeIa) is deemed to be the most robust one and therefore constitutes our baseline dataset combination, which we collectively refer to as ${\cal B}$. The combination of ${\cal B}$ and the \textit{Ly$\alpha$} BAO measurements is instead referred to as ${\cal D}$. Overall, the dataset combinations we consider are ${\cal B}$, ${\cal D}$, and ${\cal D}$+\textit{R21}.

An important clarification is in order for what concerns the adopted datasets. The reader will in fact notice that these are not the latest available ones, neither on the BAO side (where the more recent eBOSS measurements are available~\cite{eBOSS:2020yzd} and, even more recently, the DESI ones~\cite{DESI:2024mwx}), nor on the SNeIa side (where the more recent \textit{PantheonPlus} measurements are available~\cite{Scolnic:2023mrv}) whereas, for what concerns the local $H_0$ prior, we are not using the latest SH0ES measurement, although this choice is somewhat arbitrary given the deluge of available local measurements which mostly cluster between $71$ and $74\,{\text{km/s/Mpc}}$. Nevertheless, given that ours is just a proof-of-principle case study (which, furthermore, does not claim a solution to the Hubble tension), we believe this choice of datasets is not problematic, and that upgrading to the latest datasets would not alter our overall conclusions. More importantly, the adopted datasets also allow for a more fair comparison to earlier results on varying electron mass models~\cite{Sekiguchi:2020teg,Schoneberg:2021qvd}.

We obtain theoretical predictions for cosmological observables within the above models using a modified version of the Boltzmann solver \texttt{CAMB}~\cite{Lewis:1999bs}. In particular, the effects of a varying electron mass are implemented by modifying the \texttt{Recfast} module called by \texttt{CAMB}~\cite{Seager:1999bc}.\footnote{Our set-up is very similar to that of the latest \textit{Planck} analyses~\cite{Planck:2018vyg}. The recombination code difference between \texttt{Recfast++}, which is the modified version of \texttt{Recfast}, and \texttt{CosmoRec} is discussed in Ref.~\cite{Hart:2017ndk} and is completely negligible at the current level of precision of observational measurements. We also note that papers using the \texttt{CLASS} Boltzmann solver obtain results which are consistent with the two previous works~\cite{Schoneberg:2021qvd,Khalife:2023qbu} .} To sample the posterior distributions for the parameters of the models in question, we make use of Monte Carlo Markov Chain (MCMC) methods, via the cosmological sampler \texttt{CosmoMC}~\cite{Lewis:2002ah}. The convergence of the generated MCMC chains is assessed via the Gelman-Rubin parameter $R-1$~\cite{Gelman:1992zz}, and we require $R-1<0.03$ for our chains to be considered converged. Plots of the parameter posterior distributions are produced by making use of the \texttt{GetDist} package~\cite{Lewis:2019xzd}. Finally, to conduct model comparison, we calculate the Bayesian evidence for each model and estimate the corresponding relative Bayes factors, normalized to a baseline $\Lambda$CDM scenario. We use the \texttt{MCEvidence} package, which is publicly available~\cite{Heavens:2017hkr,Heavens:2017afc}\footnote{The \texttt{MCEvidence} package can be accessed at \url{https://github.com/yabebalFantaye/MCEvidence}.}. We adopt the convention that a negative Bayes factor indicates a preference for the given model over $\Lambda$CDM. To interpret the results, we refer to the revised Jeffrey's scale by Trotta~\cite{Kass:1995loi,Trotta:2008qt}, classifying evidence as inconclusive if $0 \leq | \ln B_{ij}| < 1$, weak if $1 \leq | \ln B_{ij}| < 2.5$, moderate if $2.5 \leq | \ln B_{ij}| < 5$, strong if $5 \leq | \ln B_{ij}| < 10$, and very strong if $| \ln B_{ij}| \geq 10$.

\section{Results}
\label{sec:results}

\begin{table*}[!t]
\[\begin{tabular}{|l?c|c|c|}
\hline \hline
\backslashbox{\textbf{Parameter}}{\textbf{Model}} & \ensuremath{\Lambda}CDM &  \ensuremath{\Lambda}CDM+\ensuremath{m_{e}}+\ensuremath{\Omega_{k}}  &  \ensuremath{\Lambda_{s}}CDM+\ensuremath{m_{e}}\\
\hline \hline  {\boldmath\ensuremath{\Omega_{b}h^{2}}}  & \ensuremath{0.02244\pm0.00013} & \ensuremath{0.02288_{-0.00050}^{+0.00045}} & \ensuremath{0.02209_{-0.00020}^{+0.00037}}\\
 {\boldmath\ensuremath{\Omega_{c}h^{2}}}  & \ensuremath{0.11918\pm0.00094} & \ensuremath{0.1216\pm0.0022} & \ensuremath{0.1177_{-0.0023}^{+0.0029}}\\
 {\boldmath\ensuremath{\Omega_{K}}}  & - & \ensuremath{-0.0041\pm0.0049} & -\\
 {\boldmath\ensuremath{z_{\dagger}}}  & - & - & \ensuremath{2.32_{-0.83}^{+0.60}}\\
 {\boldmath\ensuremath{m_{e}/m_{e0}}}  & - & \ensuremath{1.019\pm0.018} & \ensuremath{0.986_{-0.009}^{+0.017}}\\
 \hline \hline
 \ensuremath{H_{0}\,[${\text{km}}/{\text{s}}/{\text{Mpc}}$]}  & \ensuremath{67.74\pm0.42} & \ensuremath{69.7\pm1.7} & \ensuremath{67.1_{-1.3}^{+1.5}}\\
 \ensuremath{\Omega_{m}}  & \ensuremath{0.3101\pm0.0056} & \ensuremath{0.299\pm0.011} & \ensuremath{0.3117_{-0.0094}^{+0.0084}}\\
 \ensuremath{\sigma_{8}}  & \ensuremath{0.8099_{-0.0065}^{+0.0057}} & \ensuremath{0.825\pm0.014} & \ensuremath{0.803\pm0.013}\\
 \ensuremath{r_d\,[${\text{Mpc}}$]}  & \ensuremath{147.14\pm0.25} & \ensuremath{144.45\pm2.53} & \ensuremath{149.33\pm2.33}\\
\hline \hline
\end{tabular}\]
\caption{68\%~C.L. intervals on selected cosmological parameters indicated in the first column, as inferred within the cosmological models indicated in the first row, in light of the $\mathcal{B}$ dataset combination which does not include \textit{Ly$\alpha$} BAO measurements. The first set of parameters in bold are explicitly varied, whereas the final set of parameters not in bold are derived.}
\label{tab:b}
\end{table*}

\begin{table*}[!t]
\[\begin{tabular}{|l?c|c|c|c|}
\hline \hline
\backslashbox{\textbf{Parameter}}{\textbf{Model}} &  \ensuremath{\Lambda}CDM  &  \ensuremath{\Lambda_{s}}CDM  &  \ensuremath{\Lambda}CDM+\ensuremath{m_{e}}  &  \ensuremath{\Lambda_{s}}CDM+\ensuremath{m_{e}}\\
\hline \hline {\boldmath\ensuremath{\Omega_{b}h^{2}}}  &  \ensuremath{0.02244\pm0.00014}  &  \ensuremath{0.02236\pm0.00014}  &  \ensuremath{0.02253\pm0.00016} & \ensuremath{0.02207_{-0.00024}^{+0.00034}}\\
 {\boldmath\ensuremath{\Omega_{c}h^{2}}}  &  \ensuremath{0.11908\pm0.00093}  &  \ensuremath{0.1201\pm0.0010}  &  \ensuremath{0.1206\pm0.0019} & \ensuremath{0.1175_{-0.0023}^{+0.0027}}\\
 {\boldmath\ensuremath{m_{e}/m_{e0}}}  &  -  &  -  &  \ensuremath{1.0061\pm0.0065} & \ensuremath{0.985_{-0.010}^{+0.015}}\\
 {\boldmath\ensuremath{z_{\dagger}}}  &  -  &  \ensuremath{>2.60 \quad (>2.16)}  &  - & \ensuremath{2.19_{-0.79}^{+0.34}}\\
 \hline \hline
 \ensuremath{H_{0}\,[${\text{km}}/{\text{s}}/{\text{Mpc}}$]} &  \ensuremath{67.78\pm0.42}  &  \ensuremath{68.64_{-0.61}^{+0.50}}  &  \ensuremath{68.8\pm1.1} & \ensuremath{67.1_{-1.3}^{+1.5}}\\
 \ensuremath{\Omega_{m}}  &  \ensuremath{0.3095\pm0.0056}  &  \ensuremath{0.3038\pm0.0060}  &  \ensuremath{0.3042\pm0.0079} & \ensuremath{0.3114\pm0.0090}\\
 \ensuremath{\sigma_{8}}  &  \ensuremath{0.8100\pm0.0061}  &  \ensuremath{0.8151\pm0.0064}  &  \ensuremath{0.818\pm0.010} & \ensuremath{0.802\pm0.013}\\
 \ensuremath{r_d\,[${\text{Mpc}}$]}  & \ensuremath{147.26\pm0.23} & \ensuremath{147.08\pm0.24} & \ensuremath{146.23\pm1.13} & \ensuremath{149.53\pm2.22}\\
\hline \hline
\ensuremath{ln \mathcal B_{ij}} &  -  & \ensuremath{0.33} & \ensuremath{2.60} & \ensuremath{2.42}\\
\hline
\end{tabular}\]
\[\begin{tabular}{|l?c|c|c|c|}
\hline \hline
\backslashbox{\textbf{Parameter}}{\textbf{Model}}  &  \ensuremath{\Lambda}CDM+\ensuremath{\Omega_{k}}  &  \ensuremath{\Lambda_{s}}CDM+\ensuremath{\Omega_{k}}  &  \ensuremath{\Lambda}CDM+\ensuremath{m_{e}}+\ensuremath{\Omega_{k}}  & \ensuremath{\Lambda_{s}}CDM+\ensuremath{m_{e}}+\ensuremath{\Omega_{k}}\\
\hline \hline {\boldmath\ensuremath{\Omega_{b}h^{2}}}  &  \ensuremath{0.02242\pm0.00015}  &  \ensuremath{0.02244\pm0.00015}  &  \ensuremath{0.02300\pm0.00046} & \ensuremath{0.02236\pm0.00062}\\
 {\boldmath\ensuremath{\Omega_{c}h^{2}}}  &  \ensuremath{0.1194\pm0.0014}  &  \ensuremath{0.1190\pm0.0013}  &  \ensuremath{0.1219\pm0.0022} & \ensuremath{0.1185_{-0.0028}^{+0.0032}}\\
 {\boldmath\ensuremath{\Omega_{K}}}  &  \ensuremath{0.0006\pm0.0019}  &  \ensuremath{-0.0028_{-0.0022}^{+0.0026}}  &  \ensuremath{-0.0050\pm0.0046} & \ensuremath{-0.0022\pm0.0053}\\
 {\boldmath\ensuremath{z_{\dagger}}}  &  -  &  \ensuremath{2.50_{-0.65}^{+0.42}}  &  - & \ensuremath{2.41\pm0.61}\\
 {\boldmath\ensuremath{m_{e}/m_{e0}}}  &  -  &  -  &  \ensuremath{1.023\pm0.017} & \ensuremath{0.996_{-0.023}^{+0.026}}\\
 \hline \hline
 \ensuremath{H_{0}\,[${\text{km}}/{\text{s}}/{\text{Mpc}}$]}  &  \ensuremath{67.92\pm0.63}  &  \ensuremath{68.29\pm0.67}  &  \ensuremath{70.2\pm1.7} & \ensuremath{68.0\pm2.1}\\
 \ensuremath{\Omega_{m}}  &  \ensuremath{0.3089\pm0.0060}  &  \ensuremath{0.3049\pm0.0062}  &  \ensuremath{0.296\pm0.011} & \ensuremath{0.307\pm0.013}\\
 \ensuremath{\sigma_{8}}  &  \ensuremath{0.8113\pm0.0072}  &  \ensuremath{0.8119\pm0.0071}  &  \ensuremath{0.829\pm0.014} & \ensuremath{0.809\pm0.018}\\
 \ensuremath{r_d\,[${\text{Mpc}}$]}  & \ensuremath{147.20\pm0.30} & \ensuremath{147.27\pm0.29} & \ensuremath{143.88\pm2.39} & \ensuremath{147.92\pm3.65}\\
\hline \hline
\ensuremath{ln \mathcal B_{ij}} &  \ensuremath{3.66}  & \ensuremath{0.25} & \ensuremath{5.12} & \ensuremath{5.08}\\
\hline
\end{tabular}\]
\caption{As in Tab.~\ref{tab:b}, but for the $\mathcal{D}$ dataset combination which includes \textit{Ly$\alpha$} BAO measurements. We note that the lower limits reported on $z_{\dagger}$ for the $\Lambda_{\rm s}$CDM model are at 68\%~C.L. and 95\%~C.L., with the latter being the ones in brackets. The Bayes factors $\ln \mathcal{B}{ij} = \ln \mathcal{Z}_{\rm \Lambda CDM} - \ln \mathcal{Z}_{\rm \mathcal{M}}$ are calculated as the difference between the evidence for $\Lambda$CDM and the specific model $\mathcal{M}$ (for the same dataset $\mathcal{D}$). A negative value indicates a preference for the model $\mathcal{M}$ over the $\Lambda$CDM scenario.}
\label{tab:d}
\end{table*}

\begin{table*}[!t]
\[\begin{tabular}{|l?c|c|c|c|}
\hline \hline
\backslashbox{\textbf{Parameter}}{\textbf{Model}} &  \ensuremath{\Lambda}CDM  &  \ensuremath{\Lambda_{s}}CDM  &  \ensuremath{\Lambda}CDM+\ensuremath{m_{e}}  &  \ensuremath{\Lambda_{s}}CDM+\ensuremath{m_{e}}\\
\hline \hline {\boldmath\ensuremath{\Omega_{b}h^{2}}}  &  \ensuremath{0.02260\pm0.00013}  &  \ensuremath{0.02242\pm0.00014}  &  \ensuremath{0.02280\pm0.00014}  &  \ensuremath{0.02262_{-0.00015}^{+0.00017}}\\
 {\boldmath\ensuremath{\Omega_{c}h^{2}}}  &  \ensuremath{0.11753\pm0.00085}  &  \ensuremath{0.1198\pm0.0011}  &  \ensuremath{0.1233\pm0.0017}  &  \ensuremath{0.1221\pm0.0017}\\
 {\boldmath\ensuremath{m_{e}/m_{e0}}}  &  -  &  -  &  \ensuremath{1.0196\pm0.0047}  &  \ensuremath{1.0118_{-0.0051}^{+0.0062}}\\
 {\boldmath\ensuremath{z_{\dagger}}}  &  -  &  \ensuremath{2.24_{-0.36}^{+0.14}}  &  -  &  \ensuremath{>2.78 \quad (>2.18)}\\
 \hline \hline
 \ensuremath{H_{0}\,[${\text{km}}/{\text{s}}/{\text{Mpc}}$]}  &  \ensuremath{68.53\pm0.39}  &  \ensuremath{69.93\pm0.58}  &  \ensuremath{71.32\pm0.79}  &  \ensuremath{70.96\pm0.80}\\
 \ensuremath{\Omega_{m}}  &  \ensuremath{0.2998\pm0.0050}  &  \ensuremath{0.2922\pm0.0054}  &  \ensuremath{0.2884\pm0.0055}  &  \ensuremath{0.2888\pm0.0054}\\
 \ensuremath{\sigma_{8}}  &  \ensuremath{0.8082\pm0.0061}  &  \ensuremath{0.8181\pm0.0068}  &  \ensuremath{0.8343\pm0.0087}  &  \ensuremath{0.8289\pm0.0091}\\
 \ensuremath{r_d\,[${\text{Mpc}}$]}  & \ensuremath{147.49\pm0.22} & \ensuremath{147.10\pm0.25} & \ensuremath{144.04\pm0.84} & \ensuremath{145.21\pm1.00}\\
\hline \hline
\ensuremath{ln \mathcal B_{ij}} &  -  & \ensuremath{-3.94} & \ensuremath{-5.01} & \ensuremath{-2.85}\\
\hline
\end{tabular}\]
\[\begin{tabular}{|l?c|c|c|c|}
\hline \hline
\backslashbox{\textbf{Parameter}}{\textbf{Model}} &  \ensuremath{\Lambda}CDM+\ensuremath{\Omega_{k}}  &  \ensuremath{\Lambda_{s}}CDM+\ensuremath{\Omega_{k}}  &  \ensuremath{\Lambda}CDM+\ensuremath{m_{e}}+\ensuremath{\Omega_{k}}  & \ensuremath{\Lambda_{s}}CDM+\ensuremath{m_{e}}+\ensuremath{\Omega_{k}}\\
\hline \hline {\boldmath\ensuremath{\Omega_{b}h^{2}}}  &  \ensuremath{0.02240\pm0.00015}  &  \ensuremath{0.02245\pm0.00015}  &  \ensuremath{0.02354\pm0.00030}  &  \ensuremath{0.02339_{-0.00029}^{+0.00033}}\\
 {\boldmath\ensuremath{\Omega_{c}h^{2}}}  &  \ensuremath{0.1201\pm0.0014}  &  \ensuremath{0.1194\pm0.0013}  &  \ensuremath{0.1243\pm0.0017}  &  \ensuremath{0.1232\pm0.0018}\\
 {\boldmath\ensuremath{\Omega_{K}}}  &  \ensuremath{0.0042\pm0.0018}  &  \ensuremath{-0.0011_{-0.0024}^{+0.0028}}  &  \ensuremath{-0.0095\pm0.0033}  &  \ensuremath{-0.0100_{-0.0036}^{+0.0030}}\\
 {\boldmath\ensuremath{m_{e}/m_{e0}}}  &  -  &  -  &  \ensuremath{1.045\pm0.010}  &  \ensuremath{1.037_{-0.010}^{+0.012}}\\
 {\boldmath\ensuremath{z_{\dagger}}}  &  -  &  2.1\ensuremath{9_{-0.52}^{+0.18}}  &  -  &  \ensuremath{>2.72 \quad (2.07)}\\
 \hline \hline
 \ensuremath{H_{0}\,[${\text{km}}/{\text{s}}/{\text{Mpc}}$]}  &  \ensuremath{69.50\pm0.58}  &  \ensuremath{69.84\pm0.61}  &  \ensuremath{72.46\pm0.87}  &  \ensuremath{72.16\pm0.91}\\
 \ensuremath{\Omega_{m}}  &  \ensuremath{0.2963\pm0.0051}  &  \ensuremath{0.2922\pm0.0055}  &  \ensuremath{0.2830\pm0.0055}  &  \ensuremath{0.2828\pm0.0056}\\
 \ensuremath{\sigma_{8}}  &  \ensuremath{0.8173\pm0.0073}  &  \ensuremath{0.8173\pm0.0072}  &  \ensuremath{0.8448\pm0.0093}  &  \ensuremath{0.8393\pm0.0097}\\
 \ensuremath{r_d\,[${\text{Mpc}}$]}  & \ensuremath{147.04\pm0.29} & \ensuremath{147.16\pm0.29} & \ensuremath{140.86\pm1.37} & \ensuremath{141.91\pm1.54}\\
 \hline \hline
 \ensuremath{ln \mathcal B_{ij}} &  \ensuremath{1.08}  & \ensuremath{-1.25} & \ensuremath{-5.33} & \ensuremath{-3.74}\\
\hline 
\end{tabular}\]
\caption{As in Tab.~\ref{tab:b}, but for the $\mathcal{D}$+\textit{R21} dataset combination. We note that the lower limits reported on $z_{\dagger}$ for the $\Lambda_{\rm s}$CDM+$m_e$ and $\Lambda_{\rm s}$CDM+$m_e$+$\Omega_K$ models are at 68\%~C.L. and 95\%~C.L., with the latter being the ones in brackets. As usual, the Bayes factors $\ln \mathcal{B}{ij} = \ln \mathcal{Z}_{\rm \Lambda CDM} - \ln \mathcal{Z}_{\rm \mathcal{M}}$ are calculated as the difference between the evidence for $\Lambda$CDM and the specific model $\mathcal{M}$ (for the same dataset $\mathcal{D}$+\textit{R21}). A negative value indicates a preference for the model $\mathcal{M}$ over the $\Lambda$CDM scenario.}
\label{tab:dr21}
\end{table*}

\begin{figure}[!tb]
\includegraphics[width=1.0\linewidth]{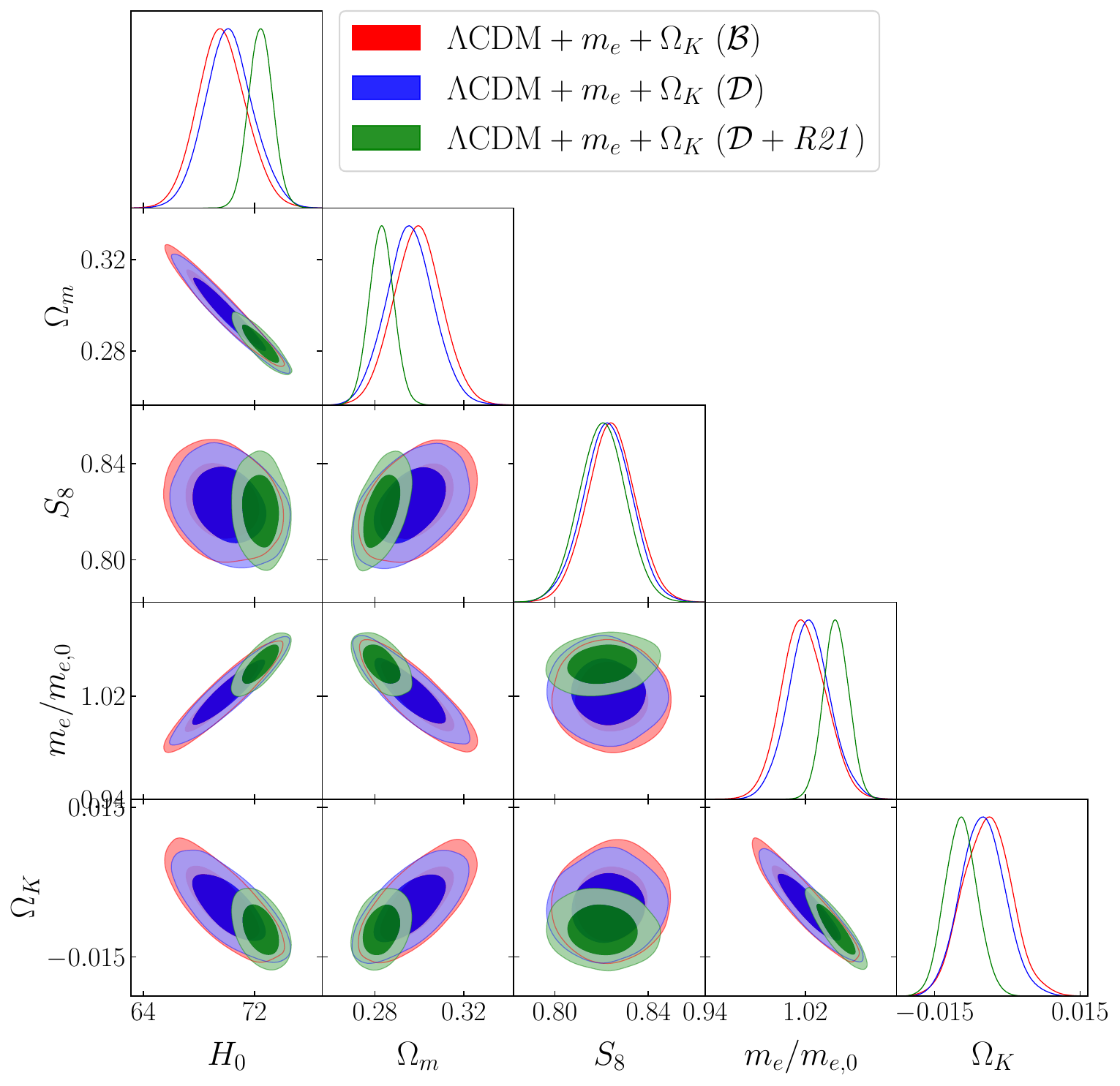}
\caption{Triangular plot showing 2D joint and 1D marginalized posterior probability distributions for a selection of parameters within the 8-parameter $\Lambda$CDM+$m_e$+$\Omega_K$ model: the Hubble constant $H_0$ (in $\text{km/s/Mpc}$), the matter density parameter $\Omega_m$, the clustering parameter $S_8$, the ratio of the electron mass at recombination to its value today $m_e/m_{e,0}$, and the curvature parameter $\Omega_K$. These distributions have been obtained in light of three different dataset combinations: the baseline combination ${\cal B}$ (red curves), the combination ${\cal D}$ (blue curves) which also includes Lyman-$\alpha$ BAO measurements, and the combination of ${\cal D}$ and \textit{R21} (green curves).}
\label{fig:me_OmegaK}
\end{figure}

\begin{figure}[!tb]
\includegraphics[width=1.0\linewidth]{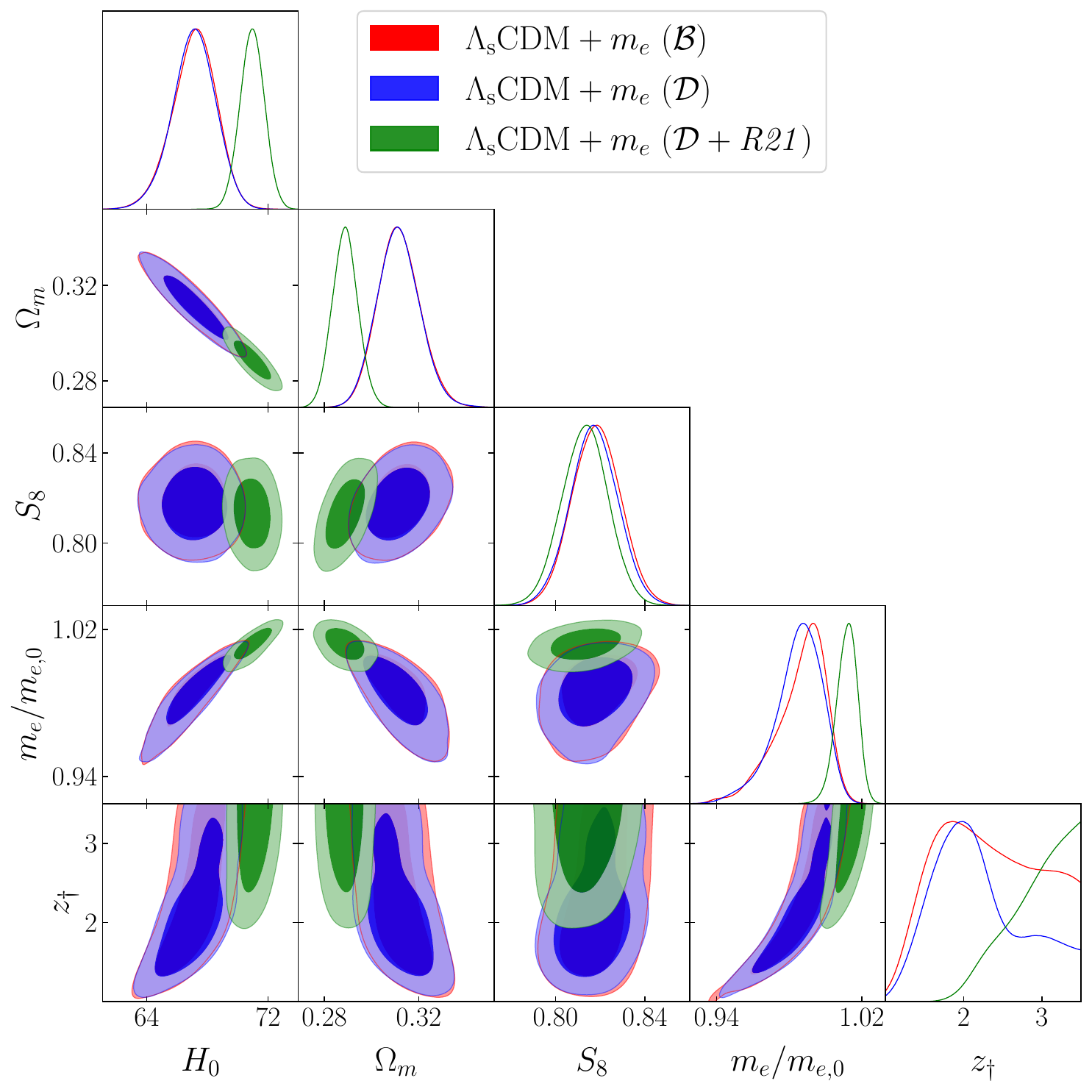}
\caption{As in Fig.~\ref{fig:me_LsCDM}, but focusing on the 8-parameter $\Lambda_{\rm s}$CDM+$m_e$ model, and reporting constraints on the AdS-to-dS transition redshift $z_{\dagger}$ instead of the curvature parameter $\Omega_K$.}
\label{fig:me_LsCDM}
\end{figure}

\begin{figure}[!tb]
\includegraphics[width=1.0\linewidth]{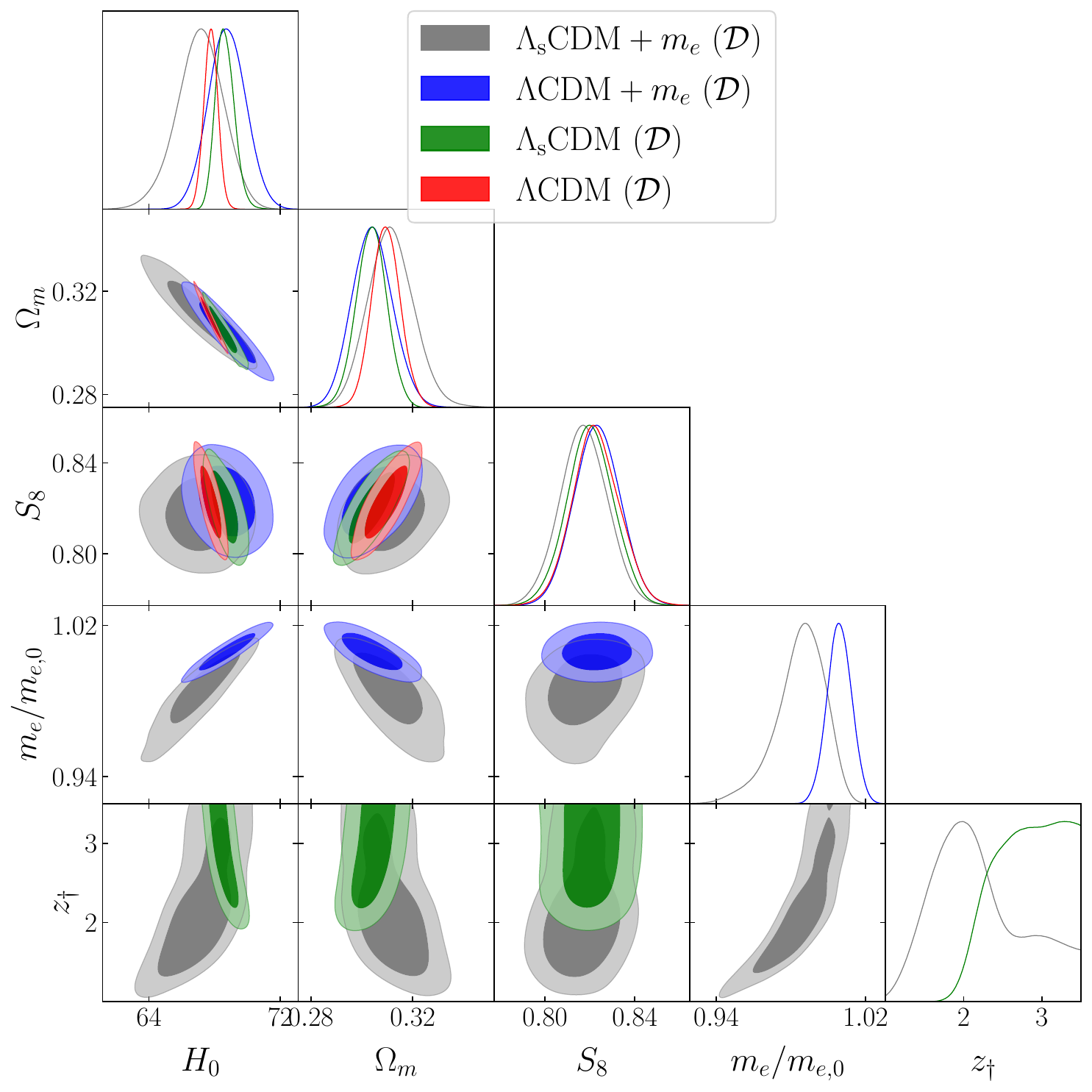}
\caption{Triangular plot comparing 2D joint and 1D marginalized posterior probability distributions for a selection of parameters across 4 spatially flat models discussed in this paper: the 6-parameter $\Lambda$CDM model (red curves), the 7-parameter $\Lambda_{\rm s}$CDM model (green curves), the 7-parameter $\Lambda$CDM+$m_e$ model (blue curves), and the 8-parameter $\Lambda_{\rm s}$CDM+$m_e$ model (grey curves). As in Fig.~\ref{fig:me_LsCDM}, we report constraints on $H_0$ (in $\text{km/s/Mpc}$), $\Omega_m$, $S_8$, $m_e/m_{e,0}$ (obviously only for the $\Lambda$CDM$+m_e$ and $\Lambda_{\rm s}$CDM$+m_e$ models), and $z_{\dagger}$ (obviously only for the $\Lambda_{\rm s}$CDM and $\Lambda_{\rm s}$CDM$+m_e$ models). These distributions have been obtained in light of the ${\cal D}$ dataset combination which also includes Lyman-$\alpha$ BAO measurements.}
\label{fig:comparison_noOmegaK_noR21}
\end{figure}

\begin{figure}[!tb]
\includegraphics[width=1.0\linewidth]{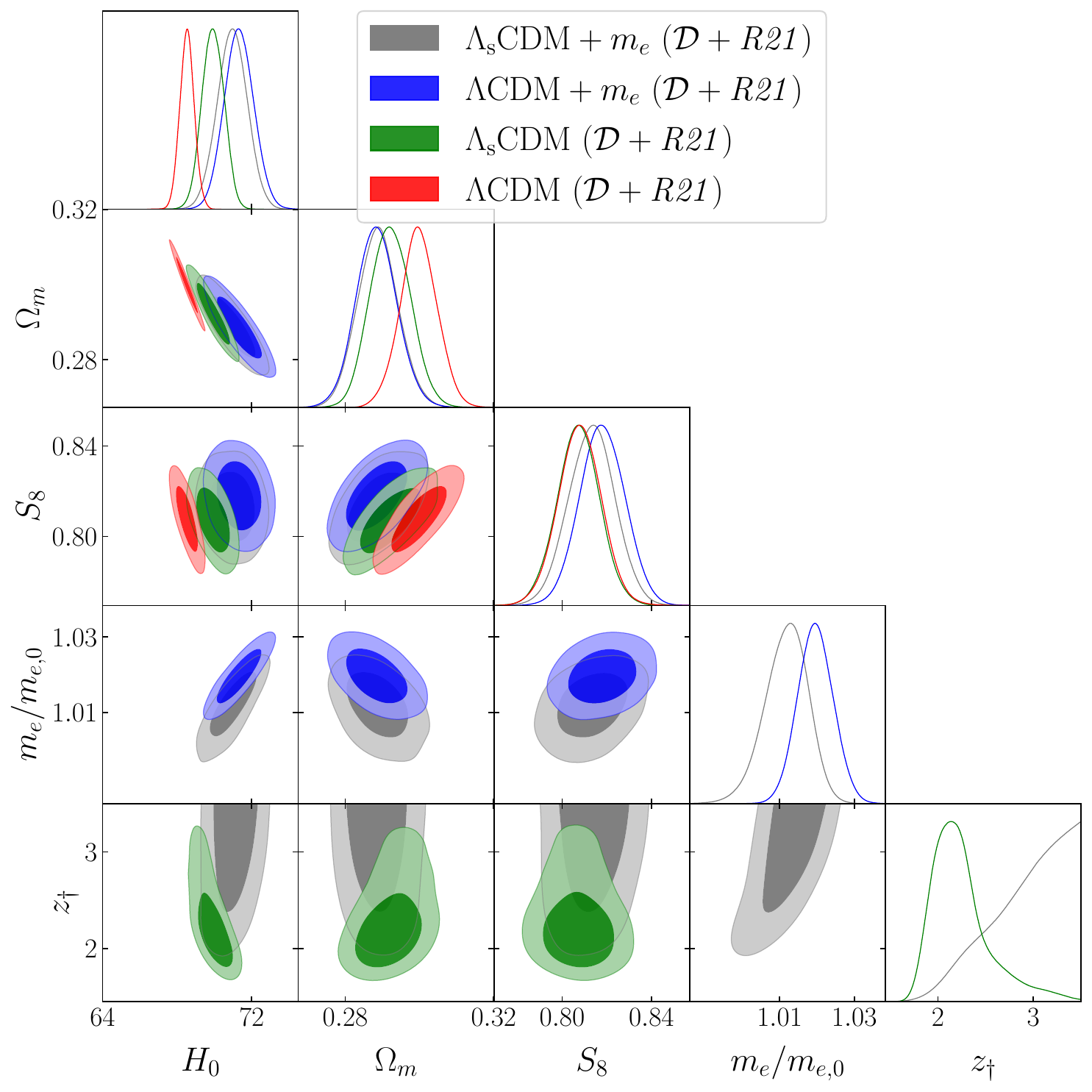}
\caption{As in Fig.~\ref{fig:comparison_noOmegaK_noR21}, but considering the combination of the ${\cal D}$ and \textit{R21} datasets.}
\label{fig:comparison_noOmegaK_R21}
\end{figure}

\begin{figure}[!tb]
\includegraphics[width=1.0\linewidth]{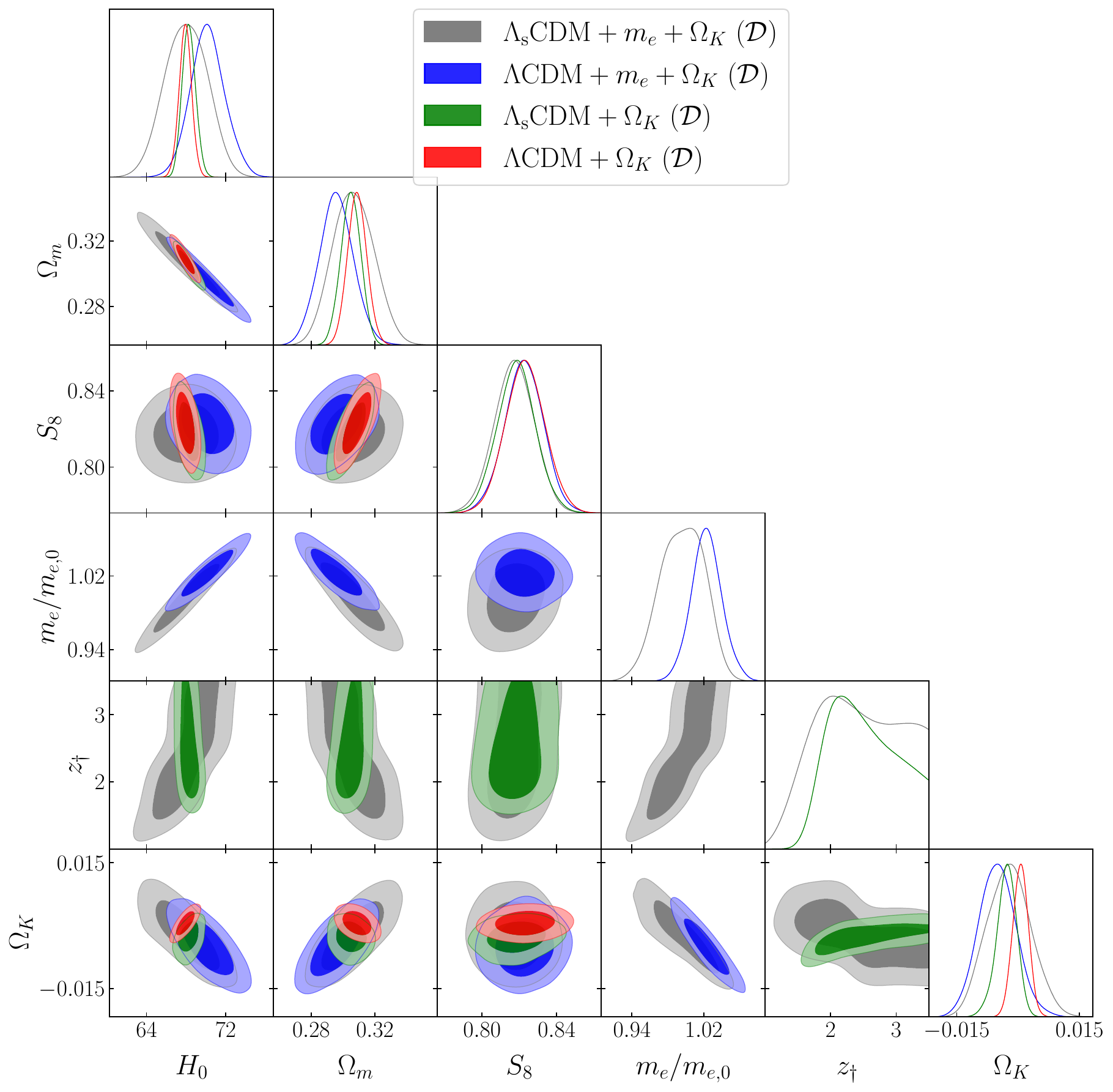}
\caption{As in Fig.~\ref{fig:comparison_noOmegaK_noR21}, but considering 4 models with varying spatial curvature: the 7-parameter $\Lambda$CDM+$\Omega_K$ model (red curves), the 8-parameter $\Lambda_{\rm s}$CDM+$\Omega_K$ model (green curves), the 8-parameter $\Lambda$CDM+$m_e$+$\Omega_K$ model (blue curves), and the 9-parameter $\Lambda_{\rm s}$CDM+$m_e$+$\Omega_K$ model (grey curves). In addition, we also report constraints on the curvature parameter $\Omega_K$.}
\label{fig:comparison_OmegaK_noR21}
\end{figure}

\begin{figure}[!tb]
\includegraphics[width=1.0\linewidth]{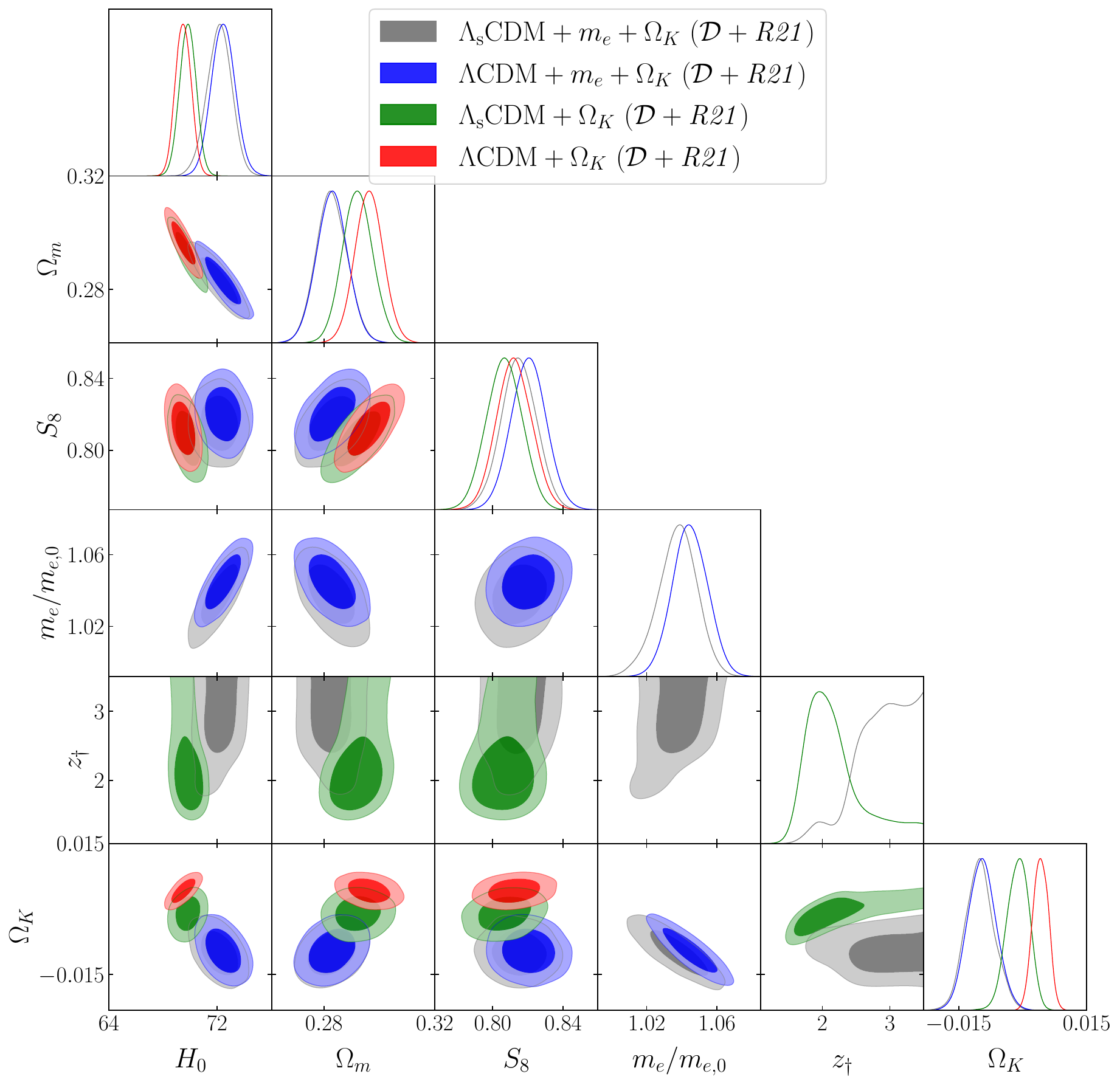}
\caption{As in Fig.~\ref{fig:comparison_OmegaK_noR21}, but considering the combination of the ${\cal D}$ and \textit{R21} datasets.}
\label{fig:comparison_OmegaK_R21}
\end{figure}

We now study the cosmological models in question in light of the dataset combinations discussed previously. Visual summaries of our results are shown in Figs.~\ref{fig:me_OmegaK},~\ref{fig:me_LsCDM},~\ref{fig:comparison_noOmegaK_noR21},~\ref{fig:comparison_noOmegaK_R21},~\ref{fig:comparison_OmegaK_noR21}, and~\ref{fig:comparison_OmegaK_R21}. On the other hand, we report 68\%~confidence level (C.L.) intervals for selected cosmological parameters in Tabs.~\ref{tab:b},~\ref{tab:d}, and~\ref{tab:dr21} for the $\mathcal{B}$, $\mathcal{D}$, and $\mathcal{D}$+\textit{R21} dataset combinations respectively. For cases where only upper/lower limits are reported in the Tables, we report both 68\%~C.L. and 95\%~C.L. upper/lower limits (the latter in brackets).

We begin by checking our results for the 8-parameter $\Lambda$CDM+$m_e$+$\Omega_K$ model, studied previously in Ref.~\cite{Sekiguchi:2020teg}. When considering the ${\cal B}$ dataset combination, we infer $H_0=(69.7 \pm 1.7)\,\text{km/s/Mpc}$, whereas we obtain $H_0=(70.2 \pm 1.7)\,\text{km/s/Mpc}$ when considering the ${\cal D}$ dataset combination -- both are significantly lower than the $H_0=72.3^{+2.7}_{-2.8}\,\text{km/s/Mpc}$ obtained in Ref.~\cite{Sekiguchi:2020teg}, considerably reducing the ability of the model to alleviate the Hubble tension. We believe the origin of this discrepancy is to be sought in the adopted data: the ${\cal B}$ dataset combination we have used also includes the CMB lensing power spectrum, which to the best of our understanding has not been adopted in Ref.~\cite{Sekiguchi:2020teg}. CMB lensing data is known to remove the preference for $\Omega_K<0$ present when using \textit{Planck} primary CMB data alone (see the earlier footnote~6), and indeed we find $\Omega_K=-0.0041 \pm 0.0049$, consistent with a spatially flat Universe within better than $1\sigma$. Given the crucial role played by $\Omega_K<0$ in the tension-solving ability of the $\Lambda$CDM+$m_e$+$\Omega_K$ model, it is clear that pushing $\Omega_K$ towards less negative values (i.e.\ closer to $\Omega_K=0$) will reduce $H_0$, while also reducing $m_e/m_{e,0}$ -- in fact, we find $m_e/m_{e,0}=1.019 \pm 0.018$, consistent with the standard value at $\approx 1\sigma$, and smaller than the $\gtrsim 5\%$ increase one would need to fully solve the Hubble tension~\cite{Sekiguchi:2020teg}.

We note that the analysis of Ref.~\cite{Schoneberg:2021qvd} made use of a similar dataset to ours (using CMB lensing, and without using \textit{Ly$\alpha$}) to infer $H_0=(69.3 \pm 2.1)\,\text{km/s/Mpc}$, in very good agreement with our findings. This comparison reinforces our interpretation that it is the CMB lensing dataset that is the main driver of the differences with respect to the earlier results of Ref.~\cite{Sekiguchi:2020teg}. On the other hand, we do not believe the choice of recombination code (\texttt{Recfast} versus \texttt{HyRec}) or sampling algorithm (Metropolis-Hastings versus nested sampling) plays a major role, in the latter case because of the high level of convergence of our chains coupled with the unimodal nature of the posteriors. 2D joint and 1D marginalized posterior probability distributions for selected parameters within the $\Lambda$CDM+$m_e$+$\Omega_K$ model, for all three dataset combinations ${\cal B}$, ${\cal D}$, and ${\cal D}$+\textit{R21}, are shown in Fig.~\ref{fig:me_OmegaK}.

For the 7-parameter $\Lambda_{\rm s}$CDM model, we essentially reproduce the earlier results of Ref.~\cite{Akarsu:2022typ}, 
even if we use a wider prior on $z_{\dagger} \in [1;3.5]$ compared to Ref.~\cite{Akarsu:2022typ}, and a slightly different combination of BAO data. The reason why widening the $z_{\dagger}$ prior leads to negligible shifts in other cosmological parameters, especially in $H_0$ which is the main parameter of interest, is that the extent to which the model approaches $\Lambda$CDM in the $z_{\dagger}\to\infty$ limit is actually non-linear, and this can easily be appreciated from the $H_0$-$z_{\dagger}$ relation shown in Fig.~8 of Ref.~\cite{Akarsu:2021fol}. Stated differently, if $z_{\dagger} \gtrsim 3$, the transition happens at such a high redshift that the cosmological constant is already completely subdominant with respect to the matter component, so the model is already basically $\Lambda$CDM (i.e.\ $z_{\dagger} \gtrsim 3$ is already a good approximation to $z_{\dagger}\to\infty$). Therefore, for what concerns cosmological parameter estimation, a prior range $z_{\dagger} \in [1;3.5]$ is sufficiently large to basically encompass the $\Lambda$CDM limit.

We now move on to examining the impact of combining one or more of the ingredients discussed previously, focusing first on spatially flat models. We begin by considering the simplest combination of all the models presented previously, namely the 8-parameter $\Lambda_{\rm s}$CDM+$m_e$ model. We show 2D joint and 1D marginalized posterior probability distributions for selected parameters within this model in Fig.~\ref{fig:me_LsCDM}, again for all three dataset combinations: ${\cal B}$, ${\cal D}$, and ${\cal D}$+\textit{R21}. A comparison between the $\Lambda$CDM (red curves), $\Lambda_{\rm s}$CDM (green curves), $\Lambda$CDM+$m_e$ (blue curves), and $\Lambda_{\rm s}$CDM (grey curves) models is instead shown in Fig.~\ref{fig:comparison_noOmegaK_noR21} for the ${\cal D}$ dataset combination, and similarly in Fig.~\ref{fig:comparison_noOmegaK_R21} for the ${\cal D}$+\textit{R21} dataset combination.

The result of this first case study is somewhat surprising. Despite both the $\Lambda_{\rm s}$CDM and $\Lambda$CDM$+m_e$ models faring better than the $\Lambda$CDM model in terms of alleviating the Hubble tension, reducing it to $3.9\sigma$ and $2.8\sigma$ respectively (when using the ${\cal D}$ dataset combination), the combined $\Lambda_{\rm s}$CDM$+m_e$ model actually fares worse than both of them in terms of the inferred central value of $H_0$. In fact, within $\Lambda_{\rm s}$CDM+$m_e$ we infer $H_0=67.1^{+1.5}_{-1.3}\,\text{km/s/Mpc}$ from both the ${\cal B}$ and ${\cal D}$ dataset combinations, bringing the tension to the $3.3\sigma$ level. We observe that the central value of $H_0$ has actually moved below the value inferred within $\Lambda$CDM, and the Hubble tension is only formally alleviated by virtue of the substantially larger uncertainties, which are a factor of $\sim 3-4$ larger with respect to those within $\Lambda$CDM (this prevents us from considering such a model a truly satisfactory resolution). Similar considerations hold when considering the ${\cal D}$+\textit{R21} dataset combination, see Fig.~\ref{fig:comparison_noOmegaK_R21}.

The above is clearly not a satisfactory situation: to put it plainly, we are finding that the whole is less/worse than the sum of its parts with regards to the ability to solve the Hubble tension (which, just to be clear, is on its own not a good metric for model performance~\cite{Cortes:2023dij}). Part of the reason behind this has to do with the inferred value of $m_e/m_{e,0}$. Recall, as per our discussion in Sec.~\ref{subsec:lcdmme}, that $m_e/m_{e,0}>1$ is required in order for the Hubble tension to be alleviated, since the geometrical degeneracy results in a positive correlation between $m_e/m_{e,0}$ and $H_0$. Physically, this is because raising $m_e/m_{e,0}$ moves recombination to an earlier time, therefore reducing the sound horizon and requiring a higher value of $H_0$ to maintain the acoustic angular scale fixed. With these considerations in mind, within the $\Lambda$CDM+$m_e$ model we find $m_e/m_{e,0}=1.019 \pm 0.018$ and $m_e/m_{e,0}=1.0061 \pm 0.0065$ from the ${\cal B}$ and ${\cal D}$ datasets respectively. Both values are too small to fully solve the Hubble tension (which would require $m_e/m_{e,0} \gtrsim 1.05$), although they move in the right direction. On the other hand, within the $\Lambda_{\rm s}$CDM+$m_e$ model we infer $m_e/m_{e,0}=0.986^{+0.017}_{-0.009}$ and $m_e/m_{e,0}=0.985^{+0.015}_{-0.010}$, both of which actually move in the wrong direction to help with the Hubble tension: this explains why, in spite of the larger uncertainties (formally alleviating the Hubble tension), the central value of $H_0$ has actually shifted to lower values relative to $\Lambda$CDM.

From Fig.~\ref{fig:me_LsCDM}, Fig.~\ref{fig:comparison_noOmegaK_noR21}, and Fig.~\ref{fig:comparison_noOmegaK_R21} we observe that the shift of $m_e/m_{e,0}$ towards $m_e/m_{e,0}<1$ appears to be driven, at least in part, by the strong $m_e$-$\Omega_m$ negative correlation, which is further exacerbated within the $\Lambda_{\rm s}$CDM+$m_e$ model (compare e.g.\ grey versus blue curves in Fig.~\ref{fig:comparison_noOmegaK_noR21} and Fig.~\ref{fig:comparison_noOmegaK_R21}). There are several reasons underlying this degeneracy. Firstly, as already explained earlier, $m_e$ and $H_0$ are positively correlated due to the geometrical degeneracy, and thereby the necessity to keep $\theta_s$ fixed. However, viable early-time solutions typically cannot lead to substantial variations in the redshift of matter-radiation equality, implying that $\Omega_mh^2$ should be roughly constant.
Therefore, one expects a negative correlation between $\Omega_m$ and $H_0$, roughly (but not exactly) along the $\Omega_m \propto h^{-2}$ direction of parameter space. It is easy to see that this degeneracy direction well approximates the one spanned by the joint $\Omega_m$-$H_0$ posteriors. It is also worth noting that a completely identical behavior has been observed in Ref.~\cite{Lee:2022gzh}, which carried out a data-driven, non-parametric, Fisher-bias reconstruction of modifications to recombination required to solve the Hubble tension, focusing also on $m_e(z)$: see in particular Fig.~2 of the paper. The combination of the positive $m_e$-$H_0$ correlation due to the geometrical degeneracy and the negative $\Omega_m$-$H_0$ correlation driven by the redshift of matter-radiation equality leads to $m_e$ inheriting a negative correlation with $\Omega_m$ as observed in our analyses both within the $\Lambda$CDM+$m_e$ and $\Lambda_{\rm s}$CDM+$m_e$ models. Obviously BAO, SNeIa, and CMB lensing (once combined with primary CMB observations) do not tolerate too low values of $\Omega_m$, and this substantially limits the ability of these models to solve the Hubble tension.

As we already noted earlier, the $m_e$-$\Omega_m$ correlation is exacerbated when the DE sector moves from the positive cosmological constant ($\Lambda$CDM+$m_e$) to the sign-switching one ($\Lambda_{\rm s}$CDM+$m_e$). This point is worthy of further exploration. Let us begin by examining Fig.~\ref{fig:comparison_noOmegaK_noR21}, and in particular the green contours for the $\Lambda_{\rm s}$ model. Here we observe a negative correlation between $z_{\dagger}$ and $H_0$, and correspondingly a positive correlation between $z_{\dagger}$ and $\Omega_m$. Both have been reported in previous literature~\cite{Akarsu:2021fol,Akarsu:2022typ,Akarsu:2023mfb}, and can easily be understood by noting that the later the AdS-to-dS transition takes place (i.e.\ the lower $z_{\dagger}$), the higher $H_0$ and the lower $\Omega_m$ need to be in order to keep the distance to the CMB fixed, ensuring that the acoustic angular scale $\theta_s$ is fixed (since the sound horizon is not modified).

Remaining still on Fig.~\ref{fig:comparison_noOmegaK_noR21}, let us see what happens when moving from the green to the grey contours, i.e.\ from the $\Lambda_{\rm s}$CDM model to the $\Lambda_{\rm s}$CDM+$m_e$ one. In this case, the ``pull'' of the $m_e/m_{e,0}$ parameter partially overwhelms the effects of $\Lambda_{\rm s}$CDM. More specifically, the negative $m_e$-$\Omega_m$ correlation extensively discussed earlier is strong enough to revert the (weaker) correlations observed within $\Lambda_{\rm s}$CDM. This is quite clear comparing the green and grey contours in the bottom row of Fig.~\ref{fig:comparison_noOmegaK_noR21}: we see that once $m_e/m_{e,0}$ is allowed to vary, the negative $z_{\dagger}$-$H_0$ correlation becomes a weak positive one, and likewise the positive $z_{\dagger}$-$\Omega_m$ correlation becomes a weak negative one. Nevertheless, the fact that the $\Lambda_{\rm s}$CDM model prefers a higher value of $\Omega_m$ to begin with is still apparent even when $m_e/m_{e,0}$ is varied, and can be appreciated by comparing the blue versus grey contours in the $m_e$-$\Omega_m$ plane. There, we see that the effect of the AdS-to-dS transition is to increase the ``pull'' towards larger values of $\Omega_m$. Because of the negative $m_e$-$\Omega_m$ correlation discussed earlier, this pushes towards values of $m_e<m_{e,0}$, going in the wrong direction compared to what is required to solve the Hubble tension, and undoing the tension-solving work associated with $m_e$. Finally, the considerations we have drawn above hold for the ${\cal D}$ dataset combination. However, our arguments hold even when considering the addition of the \textit{R21} prior, as shown in Fig.~\ref{fig:comparison_noOmegaK_R21}, with the results being qualitatively very similar -- the only difference is that the \textit{R21} prior obviously pulls towards larger values of $H_0$, and correspondingly shifts the beyond-$\Lambda$CDM parameters in response to the main degeneracies at play (i.e.\ lower $z_{\dagger}$, higher $m_e$).

It is therefore clear that $\Omega_m$, and in particular its correlations with the beyond-$\Lambda$CDM parameters in question, plays a key role in driving the (in)success of the model combining $\Lambda_{\rm s}$CDM and a varying electron mass: the important role of $\Omega_m$ in the context of the Hubble tension had in fact already been emphasized earlier~\cite{Lin:2019htv,Lin:2021sfs,Sakr:2023hrl,Baryakhtar:2024rky,Schoneberg:2024ynd,Pedrotti:2024kpn,Poulin:2024ken}. Let us, however, return to the original motivation and philosophy of this work, motivated by the earlier Ref.~\cite{Vagnozzi:2023nrq}, namely that of providing a case study aimed at combining early- and late-time new physics in an attempt to address the Hubble tension. This specific case study highlights one of the potential issues raised in Ref.~\cite{Vagnozzi:2023nrq}, but otherwise not backed up by concrete examples (see discussion in Sec.~III thereof). Namely, that the early-time and late-time models in question may not combine ``in phase''/``constructively'', and may not be able to ``decouple'' their tension-solving effects. To put it differently, parameter shifts and/or parameter degeneracies induced by one of the two models may limit the tension-solving ability of the other model, and the other way around.
This is precisely what we observe here. In particular, the beyond-$\Lambda$CDM parameters for the $\Lambda_{\rm s}$CDM and $\Lambda$CDM+$m_e$ models, $z_{\dagger}$ and $m_e/m_{e,0}$ respectively, are both positively correlated with $\Omega_m$. However, the tension-solving routes in $\Omega_m$ parameter space point in opposite directions: larger $\Omega_m$ for $\Lambda_{\rm s}$CDM, and smaller $\Omega_m$ for $m_e$. The end product is that each of the two models limits the tension-solving abilities of the other rather than combining ``constructively'', resulting in the final value of $H_0$ being even lower compared to $\Lambda$CDM, and the tension only being formally alleviated due to the substantially larger error bars.

Motivated by the important role of spatial curvature in ensuring that a model with varying electron mass can fit CMB, BAO, and SNeIa data with a high $H_0$, we now take a step forward and combine all the ingredients studied so far, namely an AdS-to-dS transition, a varying electron mass, and a non-zero spatial curvature. Constraints on the resulting 9-parameter $\Lambda_{\rm s}$CDM+$m_e$+$\Omega_K$ model are given by the grey contours in Fig.~\ref{fig:comparison_OmegaK_noR21} and Fig.~\ref{fig:comparison_OmegaK_R21} for the ${\cal D}$ and ${\cal D}$+\textit{R21} dataset combinations respectively (other than considering $\Omega_K \neq 0$, these Figures are analogous to the earlier Fig.~\ref{fig:comparison_noOmegaK_noR21} and Fig.~\ref{fig:comparison_noOmegaK_R21}). For completeness, we also report constraints on the $\Lambda$CDM+$m_e$+$\Omega_K$, $\Lambda$CDM+$\Omega_K$, and $\Lambda_{\rm s}$CDM+$\Omega_K$ models in the Figures, although the latter two are not discussed in the text.

When varying the spatial curvature parameter, the only important quantitative difference is that the value of $H_0$ inferred within models where $m_e/m_{e,0}$ is varied is higher by $\Delta H_0 \sim (1.0-1.5)\,\text{km/s/Mpc}$ compared to the corresponding models where $\Omega_K=0$ (such an enhancement is instead not observed for the $\Lambda$CDM+$\Omega_K$ and $\Lambda_{\rm s}$CDM+$\Omega_K$ models related to the $\Lambda$CDM and $\Lambda_{\rm s}$CDM ones). However, we overall qualitatively recover the same (unsuccessful) features observed earlier. Namely, combining the three ingredients does not lead to an overall higher value of $H_0$, while correspondingly shifting the beyond-$\Lambda$CDM parameters in response to the main degeneracies at play, as can clearly be appreciated by comparing Fig.~\ref{fig:comparison_OmegaK_noR21} and Fig.~\ref{fig:comparison_OmegaK_R21}. The reasons behind these unsuccessful conclusions remain precisely the ones outlined earlier, in relation to the tension-solving routes in $\Omega_m$ parameter space pointing in opposite directions, leading to the AdS-to-dS transition limiting the tension-solving ability of the combined varying electron mass and non-flat Universe model, and vice versa. The inclusion of $\Omega_K$ somewhat tames these differences, but is unable to remove them, and therefore does not lead to a successful combination of early- and late-time new physics.

To be concrete, from the ${\cal D}$ dataset combination we find $H_0=(70.2 \pm 1.7)\,\text{km/s/Mpc}$ within the $\Lambda$CDM+$m_e$+$\Omega_K$ model, which decreases to $H_0=(68.0 \pm 2.1)\,\text{km/s/Mpc}$ within the $\Lambda_{\rm s}$CDM+$m_e$+$\Omega_K$ model, formally reducing the Hubble tension down to $2.1\sigma$ mostly by virtue of the huge uncertainties. For comparison, within the $\Lambda$CDM$+\Omega_K$ model and with the same dataset combination, we find $H_0=(67.9 \pm 0.6)\,\text{km/s/Mpc}$ which, apart from the uncertainty smaller by a factor of almost $4$, features a central value that is virtually identical to the $\Lambda_{\rm s}$CDM+$m_e$+$\Omega_K$ one. Qualitatively similar conclusions, albeit with higher values of $H_0$, are reached when considering the ${\cal D}$+\textit{R21} dataset combination. Finally, as already stressed earlier, the inclusion of the CMB lensing dataset plays an important role in driving the (in)success of these combinations when non-zero spatial curvature is considered, as it removes any possible preference for $\Omega_K<0$, which is part of the required ingredients for a model featuring a varying electron mass to accommodate a high $H_0$. There is, however, reason to be cautious about the inclusion of CMB lensing measurements because of the $\approx 2.5\sigma$ disagreement with primary temperature anisotropy measurements~\cite{Planck:2018vyg}. Nevertheless, we have chosen to include these measurements to be as conservative as possible in assessing the tension-solving abilities of the combined models we are considering.

For what concerns model comparison considerations, we observe that $\ln {\cal B}_{ij}$ is always positive, and hence $\Lambda$CDM is always preferred, when the \textit{R21} prior is not included (see the final rows of Tab.~\ref{tab:dr21}). The addition of the \textit{R21} prior essentially always leads to a preference for the extended models (except in the $\Lambda$CDM+$\Omega_K$ case, see the final rows of Tab.~\ref{tab:d}), however this preference is clearly artificial and driven by the \textit{R21} prior itself. We therefore choose to focus our discussion on the more conservative ${\cal D}$ dataset combination, from which we conclude that from a Bayesian evidence perspective $\Lambda$CDM is always preferred. The models which are least disfavored with respect to $\Lambda$CDM one in light of this dataset combination are the $\Lambda_s$CDM and $\Lambda_s$CDM+$\Omega_K$ ones, with $\ln {\cal B}_{ij}=0.33$ and $0.25$ respectively, in both cases indicating only an inconclusive preference for $\Lambda$CDM.

To sum up, the combination of pre- and post-recombination new physics in the form of a spatially uniform time-varying electron mass leading to earlier recombination, and a sign-switching cosmological constant leading to an AdS-to-dS transition, does not appear to show promise in the context of solving the Hubble tension. This remains true even when the spatial curvature parameter, which plays a key role in ensuring the viability of the varying electron mass piece of the model, is allowed to vary. While the case study at hand fails to realize the type of combination envisaged in Ref.~\cite{Vagnozzi:2023nrq}, our results have nevertheless taught us several valuable lessons. First of all, recalling that our case study combined two among the most successful (or rather, least unsuccessful) models on the pre- and post-recombination sides, our results clearly indicate that a successful combination thereof is clearly no easy task. Moreover, in providing an explicit example of a potential pitfall already outlined in Ref.~\cite{Vagnozzi:2023nrq} and discussed earlier, we have highlighted the particularly important role played by $\Omega_m$, and its degeneracies with the beyond-$\Lambda$CDM parameters of the models being combined.
In this sense, one of the key issues we identified is that the degeneracy-induced tension-solving routes in $\Omega_m$ parameter space point in opposite directions for the varying electron mass and AdS-to-dS transition ingredients. This aspect, coupled to the fact that $\Omega_m$ is tightly constrained via late-time datasets (even when these are not explicitly calibrated, see for instance Refs.~\cite{Lin:2019htv,Lin:2021sfs}), leads to (with some abuse of language) ``destructive interference'' between the models being combined. As a result, each model limits the tension-solving abilities of the other, with the final value of $H_0$ being even lower than the $\Lambda$CDM value. Such a potential issue was already discussed in Ref.~\cite{Vagnozzi:2023nrq}, and our work provides an explicit realization thereof.

Although this work considers a specific example, we can still hope to draw some valuable general lessons to guide future cosmological model-building and/or tension-solving activities. In an attempt to construct a successful combination of pre- and post-recombination new physics, an important preliminary step should involve an assessment of the tension-solving directions in the parameter space of the beyond-$\Lambda$CDM parameters and how, given the directions of the parameter degeneracies involved, these project onto the other standard (fundamental or derived) parameters. We expect that a necessary but not sufficient prerequisite for a successful combination of models is that the tension-solving routes in the space spanned by parameters (directly or indirectly) well constrained by observations point in the same directions, i.e.\ precisely the opposite of what we observed in our work for the case of the matter density parameter $\Omega_m$. In this sense, parameters one should pay particular attention to are $\Omega_m$ (as stressed throughout this work), $\omega_c$, as well as potentially $A_s$, $n_s$, and $\tau$. In our case, $n_s$ plays no significant role due to the fact that we have explicitly accounted for the scaling of the Thomson scattering cross-section with $m_e$, which helps maintain $\theta_d/\theta_s$ fixed, thereby eliminating the need for shifts in $n_s$ which would otherwise be required to push this ratio back to its standard value. However, this (obviously positive) feature is one which is somewhat peculiar to the varying $m_e$ model, and not shared by other models of pre-recombination new physics, which instead typically necessitate much higher values of $n_s$ — potentially as large as $\sim 1$ — to maintain a good fit to CMB data~\cite{Poulin:2018cxd,Ye:2020btb,Ye:2021nej,Jiang:2021bab,Jiang:2022uyg,Jiang:2022qlj,Poulin:2023lkg,Jiang:2023bsz,Peng:2023bik,Giare:2024akf}: we therefore expect that for these models, assessing how tension-solving directions project in $n_s$ parameter space may play as important a role as that of analogous considerations for $\Omega_m$ in our case. For what concerns $\omega_b$, we do not expect similar considerations thereon to play an important role, given how tightly it is constrained from the relative height of the acoustic peaks in the CMB, as well as from BBN considerations: there is hence very little wiggle room for variations in $\omega_b$ to play a significant role in similar discussions (see e.g.\ Refs.~\cite{Seto:2021xua,Schoneberg:2022ggi,Takahashi:2023twt} for discussions on $\omega_b$ in the context of the CMB, BBN, and the Hubble tension).

Two comments and a caveat are in order before closing. Our study, and thereby our discussion, have been centered upon the combination of a varying electron mass and a sign-switching cosmological constant leading to an AdS-to-dS transition. While we have tried to extract general lessons from our work, we caution the reader that some of these are inevitably specific to our model: complete and nuanced assessments on the viability (or not) of combinations of pre- and post-recombination new physics models will inevitably have to be done on a model-by-model basis, and different model combinations may flag other aspects, not apparent in our work, which may prevent such a combination from functioning. In addition, our discussion has been centered around the Hubble tension. Nevertheless, we have also checked that the combination of models we have studied has little impact on the $S_8$ discrepancy, which therefore remains unsolved. Finally, as already stressed in Sec.~\ref{sec:data}, we have not used the latest available cosmological datasets on the late-time data side. Nevertheless, we expect that updating the adopted datasets to the latest available ones will not qualitatively change our results, and neither should lead to major quantitative changes. That is, adopting these datasets should slightly tighten some of our constraints and may slightly change some degeneracy directions, but would not alter neither the overall (lack of) success of the specific combination of models considered, nor our general discussions on parameter degeneracies.

\section{Conclusions}
\label{sec:conclusions}

\begin{figure*}
\includegraphics[width=1.0\linewidth]{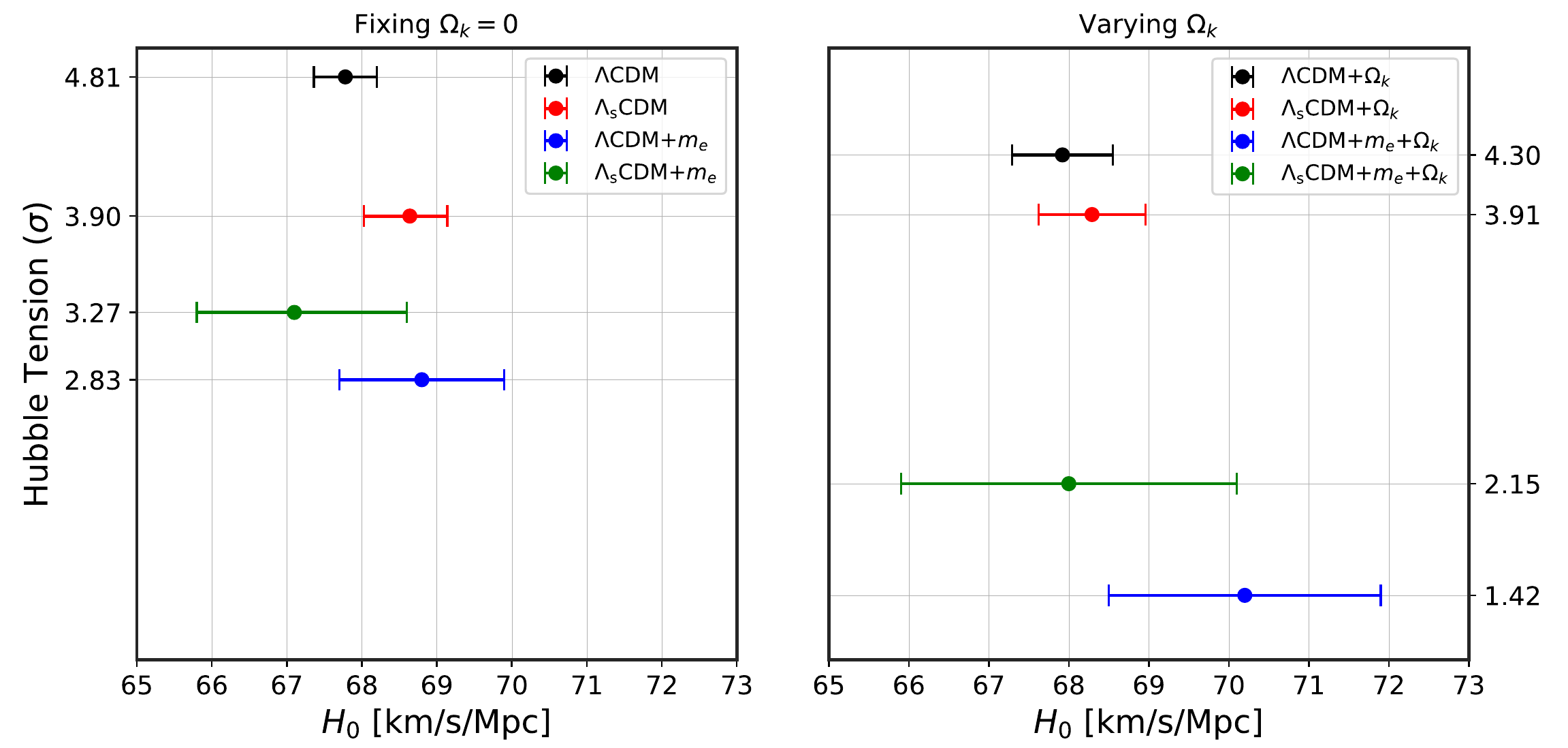}
\caption{Whisker summary plot displaying 68\%~C.L. intervals on the Hubble constant $H_0$ (on the abscissa axis, in units of ${\text{km}}/{\text{s}}/{\text{Mpc}}$) and residual level of the Hubble tension (on the ordinate axis, in units of equivalent Gaussian $\sigma$) within the 8 different models considered. The left panel focuses on the 4 spatially flat models studied -- the 6-parameter $\Lambda$CDM model (black), the 7-parameter $\Lambda_{\rm s}$CDM model (red), the 7-parameter $\Lambda$CDM+$m_e$ model (blue), and the 8-parameter $\Lambda_{\rm s}$CDM+$m_e$ model (green) -- and conversely the right panel focuses on the 4 models with varying spatial curvature studied -- the 7-parameter $\Lambda$CDM+$\Omega_K$ model (black), the 8-parameter $\Lambda_{\rm s}$CDM+$\Omega_K$ model (red), the 8-parameter $\Lambda$CDM+$m_e$+$\Omega_K$ model (blue), and the 9-parameter $\Lambda_{\rm s}$CDM+$m_e$+$\Omega_K$ model (green). These constraints have been obtained using the ${\cal D}$ dataset combination, including Lyman-$\alpha$ BAO measurements.}
\label{fig:summary}
\end{figure*}

At the time of writing, despite being one of the hottest open problems in cosmology, the Hubble tension remains unsolved after over a decade of attempts. Among the efforts in this direction based on models of new fundamental physics, the vast majority have focused on introducing new physical ingredients either before or after recombination, with the former class identified as being the least unsuccessful one due to constraints imposed from late-time observations. However, motivated by a number of independent indications, recent work by one of us has raised the possibility that solving the Hubble tension might ultimately require a combination of pre- and post-recombination new physics~\cite{Vagnozzi:2023nrq}, without however proposing any concrete combination in this sense. Driven by this possibility, in the present work we provide one of the first concrete case studies aimed at combining pre- and post-recombination new physics in an attempt to address the Hubble tension (see also Refs.~\cite{Allali:2021azp,Anchordoqui:2021gji,Khosravi:2021csn,Clark:2021hlo,Wang:2022jpo,Anchordoqui:2022gmw,Reeves:2022aoi,Yao:2023qve,daCosta:2023mow,Wang:2024dka} for earlier attempts in the same direction). In what is perhaps -- especially in hindsight -- somewhat of a na\"{i}ve attempt, we combine two models which, from each of the two sides, have \textit{individually} shown particular promise in alleviating the Hubble tension: a spatially uniform time-varying electron mass leading to earlier recombination (and eventually adding non-zero spatial curvature, given its important role in the context of this model), and a sign-switching cosmological constant (i.e., an AdS-to-dS transition, within the so-called $\Lambda_{\rm s}$CDM model).

Our study shows that no combination of the aforementioned ingredients leads to a model that successfully solves the Hubble tension. Our results are visually summarized in the whisker plot of Fig.~\ref{fig:summary}, where we see that within the $\Lambda_{\rm s}$CDM+$m_e$ and $\Lambda_{\rm s}$CDM+$m_e$+$\Omega_K$ models, the Hubble tension is only reduced to the $3.3\sigma$ and $2.2\sigma$ levels respectively. These figures (especially the latter) are, however, artifacts of very large uncertainties, which are in the latter case a factor of $\sim 3$-$4$ larger compared to the uncertainty on $H_0$ within $\Lambda$CDM (itself a consequence of the three additional parameters). It is worth noting that the mean value within the $\Lambda_{\rm s}$CDM+$m_e$+$\Omega_K$ model is actually even \textit{lower} than the $\Lambda$CDM value. We have found that this can be ascribed to a potential problem already pointed out (albeit without an explicit example) in Ref.~\cite{Vagnozzi:2023nrq}, i.e.\ that degeneracies within each of the two models being combined limit the tension-solving abilities of the other, leading to a combination where the whole is less than the sum of its parts (in terms of the value of $H_0$). We have highlighted the particularly important role of $\Omega_m$ in these considerations. 
In fact, for the two models being combined, the tension-solving routes in $\Omega_m$ parameter space point in opposite directions—the AdS-to-dS transition wants a larger $\Omega_m$ to maintain the distance to the CMB fixed, whereas the varying electron mass wants a smaller $\Omega_m$ to keep the redshift of matter-radiation equality fixed. These results highlight once more the extremely important role of $\Omega_m$ in arbitrating viable solutions to the Hubble tension. Finally, we have verified that none of the model combinations considered has a significant impact on the $S_8$ discrepancy.

Despite the overall failure to obtain a solution to the Hubble tension, we have learned some valuable lessons for future tension-solving activities attempting to construct a successful combination of pre- and post-recombination new physics, in the spirit of Ref.~\cite{Vagnozzi:2023nrq}. We have argued that for such a combination to be successful, it is important that the tension-solving routes in the space spanned by parameters well constrained by observations point in the same directions, with $\Omega_m$ playing a particularly important role in this sense. Of course, other considerations may prevent such combinations from working, but we believe the one highlighted above represents a particularly important point, which can already be preliminarily assessed even before carrying out a complete analysis within a specific combination of models. Our work opens up various interesting follow-up directions, with the most obvious one being to carry out other attempts at constructing viable combinations of pre- and post-recombination new physics, guided by the lessons we have just learned. Besides a time-varying electron mass, a particularly interesting candidate on the early-time side is, of course, early dark energy, for which we already know that degeneracies involving $\omega_c$, $n_s$, and to some extent $\Omega_m$ play a particularly important role. In terms of better controlling $\Omega_m$, interacting dark energy models at late times might offer a route of potential interest, given that they achieve a higher value of $H_0$ (albeit not fully solving the tension) while lowering $\Omega_m$ (see e.g., Refs.~\cite{DiValentino:2019ffd,DiValentino:2019jae,Wang:2024vmw}). Moreover, it could be worth considering dark scattering models~\cite{Simpson:2010vh,Kumar:2017bpv,Vagnozzi:2019kvw,BeltranJimenez:2020iyx,BeltranJimenez:2021wbq,Ferlito:2022mok,Poulin:2022sgp}, whose role in potentially helping with the $S_8$ discrepancy was pointed out in Ref.~\cite{Vagnozzi:2023nrq}. Finally, a number of more recent cosmological datasets are now available~\cite{DESI:2024mwx}; these could help strengthen our conclusions or shed further light on the ingredients required for a successful pre-plus-post-recombination new physics combination, while at the same time it may be of interest to assess the feasibility of testing such models in future cosmological data~\cite{CMB-S4:2016ple,SimonsObservatory:2018koc,SimonsObservatory:2019qwx}. We defer studies on these and related aspects to future work.

\begin{acknowledgments}
\noindent We are grateful to Vivian Poulin and Nils Sch\"{o}neberg for many helpful discussions, and to Osamu Seto for collaboration throughout various stages of the project. Y.T.\ acknowledges support from the Japan Science and Technology Agency (JST) through the SPRING (Support for Pioneering Research Initiated by the Next Generation) grant no.\ JPMJSP2119. E.\"{O}.\ acknowledges support from the T\"{u}rkiye Bilimsel ve Teknolojik Ara\c{s}t{\i}rma Kurumu (T\"{U}B\.{I}TAK, Scientific and Technological Research Council of Turkey) through the 2214/A National Graduate Scholarship Program. W.G.\ is supported by the Lancaster–Sheffield Consortium for Fundamental Physics under STFC grant ST/X000621/1.  E.D.V.\ acknowledges support from the Royal Society through a Royal Society Dorothy Hodgkin Research Fellowship. S.V.\ acknowledges support from the University of Trento and the Provincia Autonoma di Trento (PAT, Autonomous Province of Trento) through the UniTrento Internal Call for Research 2023 grant ``Searching for Dark Energy off the beaten track'' (DARKTRACK, grant agreement no.\ E63C22000500003), and from the Istituto Nazionale di Fisica Nucleare (INFN) through the Commissione Scientifica Nazionale 4 (CSN4) Iniziativa Specifica ``Quantum Fields in Gravity, Cosmology and Black Holes'' (FLAG). This publication is based upon work from the COST Action CA21136 ``Addressing observational tensions in cosmology with systematics and fundamental physics'' (CosmoVerse), supported by COST (European Cooperation in Science and Technology).
\end{acknowledgments}

\bibliography{meLsCDM}

\end{document}